\PassOptionsToPackage{numbers,square}{natbib}
\documentclass[preprint,12pt]{elsarticle}

\usepackage{amssymb}
\usepackage{amsmath}
\usepackage{graphicx}
\usepackage{multirow}
\usepackage{amsfonts}
\usepackage{amsthm}
\usepackage{mathrsfs}
\usepackage[title]{appendix}
\usepackage{xcolor}
\usepackage{textcomp}
\usepackage{manyfoot}
\usepackage{booktabs}
\usepackage{algorithm}
\usepackage{algorithmicx}
\usepackage{algpseudocode}
\usepackage{listings}
\usepackage{lineno}
\usepackage{enumitem}
\usepackage{setspace}
\usepackage[normalem]{ulem}
\usepackage[margin=2.5cm]{geometry}
\usepackage{bm}

\makeatletter
\gdef\emailauthor#1#2{%
  \stepcounter{ead}%
  \g@addto@macro\@elseads{%
    \scriptsize
    \raggedright
    \let\corref\@gobble
    \def\@@tmp{#1}%
    \eadsep
    {\ttfamily\expandafter\strip@prefix\meaning\@@tmp}
    \space(#2)%
    \def\eadsep{\unskip,\space}%
  }%
}
\makeatother

\usepackage{eso-pic}

\AddToShipoutPictureFG*{%
  \AtPageUpperLeft{%
    \put(55,-38){%
      \parbox{0.80\paperwidth}{%
        \centering
        \scriptsize
        This article has been accepted for publication in
        \textit{Mathematics and Mechanics of Complex Systems (MEMOCS)}.\\
        Please cite the paper as:\\
        M. M. Nayeban, B. E. Abali, R. Afshar, F. Fabbrocino,
        A. Misra, and L. Placidi,\\
        ``Stiffness coefficients in granular micromechanics determined
        with three-point bending experiments and digital image correlation
        for a bimodulus material,''
        \textit{Mathematics and Mechanics of Complex Systems}, in press.
      }%
    }%
  }%
}

\begin{document}

\begin{frontmatter}

\title{Stiffness coefficients in granular micromechanics determined with three-point bending experiments and digital image correlation for a bimodulus material}

\author[mn1,mn2]{Mohammad Mahdi Nayeban\corref{cor1}}
\ead{mohammad.nayeban@phd.unict.it}

\author[mn3]{Bilen Emek Abali}
\ead{b.emek.abali@angstrom.uu.se}

\author[mn3]{Reza Afshar}
\ead{reza.afshar@angstrom.uu.se}

\author[mn4]{Francesco Fabbrocino}
\ead{francesco.fabbrocino@unipegaso.it}

\author[mn5]{Anil Misra}
\ead{anmisra@fiu.edu}

\author[mn4]{Luca Placidi}
\ead{luca.placidi@unipegaso.it}

\cortext[cor1]{Corresponding author}

\affiliation[mn1]{organization={University of Catania}, country={Italy}}
\affiliation[mn2]{organization={International Telematic University Uninettuno}, country={Italy}}
\affiliation[mn3]{organization={Uppsala University}, country={Sweden}}
\affiliation[mn4]{organization={Telematic University Pegaso}, country={Italy}}
\affiliation[mn5]{organization={Florida International University}, country={USA}}

\begin{abstract}
This study presents an inverse analysis to determine the stiffness parameters of a bimodulus constitutive model formulated within a granular micromechanics framework. Full-field normal strain measurements are obtained by Digital Image Correlation (DIC) during three-point bending tests and are used as experimental input to a hybrid numerical optimization procedure. The experimental campaign is designed to ensure that the measured strain fields remain within the elastic regime.
Since the identification is formulated as a nonlinear regression problem, a physically informed initial parameter set is constructed to guide the optimization toward mechanically meaningful values. An initial estimate of the compressive stiffness is first obtained from the available manufacturer data. The tension–compression stiffness asymmetry is then estimated from the experimentally measured neutral-axis shift, which provides a macroscopic signature of bimodulus material behavior in bending. These estimates define the initial parameter set, which is subsequently refined through a hybrid optimization procedure based on the full-field comparison of the experimental and numerical normal strain distributions.
The identified parameter set enables the numerical model to reproduce more closely both the through-depth profile and the spatial distribution of the normal strain component over the selected surface region.
These results demonstrate that full-field normal strain measurements provide an effective basis for identifying the stiffness parameters of bimodulus granular micromechanics models. The largest differences between the numerical and experimental strain fields are observed near the load-application region, where localized strain variations are induced by the concentrated load.
\end{abstract}

%-----------------%

%% Keywords

\end{frontmatter}

% =================================================
\section{Introduction}

Reliable application of constitutive models requires the identification of the stiffness parameters governing the mechanical response from experimental data. This task becomes more demanding when a material exhibits different elastic responses in tension and compression, because a single elastic modulus is no longer sufficient to describe its behavior. Materials exhibiting this type of response are commonly referred to as bimodulus materials and are often characterized at the macroscopic level by distinct Young's moduli in tension and compression. Representative examples include lime mortar, natural building stones, graphite, and certain ceramics.
Accounting for this asymmetry is essential for the accurate prediction of stress and strain distributions in structural members. Its effect is particularly evident in bending, where structural members with geometrically symmetric cross-sections nevertheless exhibit a neutral-axis shift away from the section centroid when the tensile and compressive stiffnesses differ. This shift therefore provides a macroscopic signature of bimodular behavior \cite{parisi2015experimental, rohanifar2021numerical}.

Several continuum models have been proposed to describe materials with different elastic responses in tension and compression. Classical approaches originating from Ambartsumyan's bimodulus theory distinguish tensile and compressive behavior according to the signs of the principal stresses \cite{ambartsumyan1986elasticity}. Subsequent developments have introduced general computational frameworks \cite{du2016new}, energy-consistent reformulations \cite{latorre2020bi}, and inverse identification procedures \cite{ran2018gradient}. Related bimodulus formulations have also been applied to structural topology optimization and composite material design \cite{du2019structural,cai2016optimal,gaganelis2019tension,guo2023topology}.
Despite these developments, many existing formulations determine the tensile or compressive response through macroscopic switching criteria defined in stress or strain space, often requiring separate constitutive cases for different combinations of tensile and compressive loading. In such approaches, the tension/compression classification is imposed at the continuum level, while the microscopic origin of the asymmetry is not explicitly represented. Inverse identification has likewise been formulated within this macroscopic switching structure \cite{ran2018gradient}. In contrast, the granular micromechanics approach introduces the asymmetry directly at the level of orientation-dependent normal and tangent interactions, allowing the macroscopic bimodulus response to emerge from the underlying interaction mechanics.
Within this framework, each material point of the continuum is associated with a continuous distribution of pair interactions over the space of orientations. For each orientation, the interaction deformation is decomposed into a normal contribution along the interaction direction and a tangent contribution orthogonal to it. The corresponding interaction stiffnesses describe the resistance associated with these two deformation modes, and the macroscopic deformation energy is obtained by integrating their contributions over all orientations.
The tensile or compressive state is therefore not assigned to the material point as a whole, but to each interaction orientation according to the sign of the corresponding normal displacement. Consequently, a material point is generally not entirely in tension or entirely in compression in all directions. This orientation-dependent description makes it possible to introduce tension/compression asymmetry directly through the normal and tangent stiffnesses associated with each interaction orientation, rather than relying exclusively on a single macroscopic modulus ratio \cite{placidi2026revised, misra2025emergence, poorsolhjouy2021granular}. When the stiffness distribution is uniform over the orientational space, the isotropy of the reference material is preserved.

An earlier beam study \cite{nayeban2025analysis} applied this framework to bimodulus beams and investigated how tension/compression asymmetry modifies the strain and stress fields and produces a neutral-axis shift. Building on that study, the present work addresses the corresponding inverse problem, namely whether the stiffness parameters governing this asymmetry can be identified from full-field experimental data. The main objective of the present work is to identify
simultaneously the four independent parameters governing the bimodulus
interaction response from the measured full-field strain field. For this purpose, the analysis is restricted to the classical first-gradient formulation, allowing the identification procedure to remain focused on the dominant full-field strain response and on the bimodulus stiffness parameters while maintaining a sufficiently simple constitutive structure.

The granular micromechanics framework has also been extended in the literature to strain-gradient continua through the same micro-to-macro construction based on Piola's ansatz. In these formulations, additional kinematical descriptors and characteristic lengths can be introduced, and second-gradient elastic parameters have been related to classical elastic constants and to a microstructural length scale \cite{placidi2021micromechanics, pirmoradi2024anisotropic, stilz2022continuum}. Such extensions become relevant when localized deformation mechanisms or size-dependent effects need to be represented. Although they are not included in the present identification procedure, they provide a possible direction for future developments, particularly for improving the description of localized effects near the load-application region.

Experimental full-field strain measurements can be obtained by digital image correlation (DIC), an optical and non-contact image-based technique that enables the evaluation of displacement and strain fields over a selected region of interest \cite{schreier2002systematic}. When the observed surface remains approximately planar and out-of-plane motions are limited, two-dimensional DIC provides a convenient framework for strain analysis under mechanical loading. In the present study, DIC is employed in combination with three-point bending tests to obtain the full-field strain response required for the identification procedure.
The resulting DIC measurements serve as the experimental input for the inverse analysis \cite{merzkirch2021mechanical}. Full-field measurements are particularly useful when local strain distributions, strain localization, and the dependence of the extracted quantities on the selected measurement region are relevant to the mechanical characterization. Rather than relying on a single global quantity, such as the load--displacement response, the proposed approach therefore uses the spatial distribution of the normal strain field over the selected region of interest (ROI) to identify the constitutive parameters.
Similar uses of full-field measurements for model calibration and parameter identification have been reported in the literature \cite{giorgio2018wrinkling, fedele2013motion, fedele2022bayesian}. Accordingly, the identification strategy adopted here is based on the discrepancy between the measured and simulated full-field normal strain distributions over the selected ROI, rather than solely on derived quantities such as the neutral-axis position.

The identification problem requires repeated evaluations of the objective function through a coupled MATLAB--COMSOL workflow. Given the computational nature of this workflow and the lack of directly available gradient information, the optimization is carried out using a two-stage strategy that combines a derivative-free exploration phase with a subsequent local refinement phase. Purely gradient-based approaches are not straightforward, particularly during the initial stages of the search, since gradient information is not directly available from the numerical workflow. A pattern-search strategy is first adopted to explore the parameter space and identify a suitable neighborhood of the optimum. Such methods are particularly appropriate when the objective function is available only through numerical evaluations and may exhibit limited smoothness \cite{hooke1961direct,Torczon1997,LewisTorczon1999}.
Once a suitable parameter region has been identified, the solution is further refined through a local procedure based on quasi-Newton methods \cite{xu2001survey,NocedalWright2006}. In this class of methods, local curvature information is approximated to construct an efficient search direction, generally providing faster convergence in the vicinity of the optimum. Quasi-Newton methods, and in particular BFGS-type updates, are well established for this purpose in nonlinear optimization.
The present study therefore combines full-field DIC measurements, a granular micromechanics-based forward model, and a hybrid optimization strategy to identify a set of bimodulus stiffness parameters consistent with the experimentally measured bending response \cite{giorgio2022experimental}.

% % =================================================
\section{Granular micromechanics based modeling}\label{sec:Methodology}

\subsection{Kinematics and energy formulation}\label{sec:kinematics_energy}
In a two-dimensional continuum body, $\Omega$, the granular micromechanics approach
represents the elastic response at a material point $\mathbf{X}\in\Omega$ as the
aggregate effect of pairwise interactions distributed over all orientations
$\theta\in[0,2\pi]$.  
Each orientation is described by a unit vector
$\hat{\mathbf{c}}\in S^{1}$, where $S^{1}$ denotes the unit circle. A standard representation of the unit vector $\hat{\mathbf c}$ in terms of the orientation angle $\theta$ is,
\begin{equation}
\hat{c}_1=\cos(\theta),
\qquad
\hat{c}_2=\sin(\theta).
\end{equation}
For each interaction, the deformation is decomposed into a normal displacement $u_\eta$ along
$\hat{\mathbf{c}}$ and a tangent displacement $u_\tau$ orthogonal to it, governed by the
(normal and tangent) stiffnesses $k_\eta$ and $k_\tau$, respectively.
The local strain-energy density $U$ is expressed through the following orientation integral:
\begin{equation}
U(\mathbf{X})=\int_{S^{1}}
\left[
\frac{1}{2}k_\eta\,u_\eta^{2}
+
\frac{1}{2}k_\tau\,u_\tau^{2}
\right]\mathrm{d}\theta,
\qquad \forall\,\mathbf{X}\in\Omega .
\label{eq:energy_density_rev}
\end{equation}
The normal and tangent displacement measures are written in terms of the
Green--Saint-Venant strain components $G_{ij}$ and a characteristic microstructural length $L$ as
\begin{equation}
u_\eta = L\,G_{ij}\,\hat{c}_i\hat{c}_j,
\qquad
u_\tau^{2}
=
4L^{2}\,G_{ij}G_{ab}
\left(
\delta_{ia}\hat{c}_j\hat{c}_b
-
\hat{c}_i\hat{c}_j\hat{c}_a\hat{c}_b
\right),
\label{eq:micro_measures_rev}
\end{equation}
where $\delta_{ij}$ denotes the Kronecker delta and $G_{ij}$ denote the components of the Green--Saint-Venant (Lagrangian finite) strain tensor $\bm{G}$. Accordingly, $\hat{c}_i$ denotes the $i$-th component of the unit vector $\hat{\bm c}$. Repeated indices imply summation throughout this paper. The kinematics is described by the placement field $\bm{\chi}$ and the deformation gradient $\bm{F}$,

\begin{equation}
\bm{F}=\nabla\bm{\chi},
\qquad
G_{ij}=\frac{1}{2}\left(F_{k i}F_{k j}-\delta_{ij}\right),
\label{eq:F_Gij_def_rev}
\end{equation}
which is equivalent to $\bm{G}=\tfrac{1}{2}(\bm{F}^{\top}\bm{F}-\bm{I})$ in matrix form.

Substituting Eq.~\eqref{eq:micro_measures_rev} into Eq.~\eqref{eq:energy_density_rev} yields the
quadratic macroscopic energy representation
\begin{equation}
U=\frac{1}{2}\,C_{ijab}\,G_{ij}\,G_{ab},
\label{eq:compact_energy_rev}
\end{equation}
from which the components of the fourth-order stiffness tensor $\bm{C}$ are identified as
\begin{equation}
\begin{aligned}
C_{ijab} = L^2 \int_{S^1}
& k_\eta\,\hat{c}_i\hat{c}_j\hat{c}_a\hat{c}_b\,\mathrm{d}\theta \\
+\, L^2 \int_{S^1}
& k_\tau\Big[
\delta_{ia}\hat{c}_j\hat{c}_b
+\delta_{ib}\hat{c}_j\hat{c}_a
+\delta_{ja}\hat{c}_i\hat{c}_b
+\delta_{jb}\hat{c}_i\hat{c}_a
-4\hat{c}_i\hat{c}_j\hat{c}_a\hat{c}_b
\Big]\,\mathrm{d}\theta .
\end{aligned}
\label{eq:stiffness_tensor_rev}
\end{equation}

\subsection{Stiffness identification and tension/compression asymmetry}
\label{sec:asymmetry}

As a reference, for homogeneous isotropic materials with tension/compression symmetric response, the interaction stiffnesses $k_\eta$ and $k_\tau$ are taken as constants, independent of the placement field $\bm{\chi}$ and of the orientation $\theta$. Following \cite{placidi2026revised}, their expressions in terms of the effective 2D Young's modulus $Y_{2D}$ and Poisson's ratio $\nu_{2D}$ are
\begin{equation}
k_\eta = \frac{Y_{2D}}{\pi L^2 (1-\nu_{2D})},
\qquad
k_\tau = \frac{Y_{2D}(1-3\nu_{2D})}{4\pi L^2 (1-\nu_{2D})(1+\nu_{2D})},
\label{eq:isotropic_stiffness_rev}
\end{equation}
and, conversely,
\begin{equation}
\label{eq:A_general}
Y_{2D} = 2\pi L^{2} k_{\eta}
\frac{k_{\eta}+4k_{\tau}}{3k_{\eta}+4k_{\tau}},
\qquad
\nu_{2D}= \frac{k_{\eta}-4k_{\tau}}{3k_{\eta}+4k_{\tau}}.
\end{equation}
The quantities $Y_{2D}$ and $\nu_{2D}$ should be interpreted as effective parameters of the present two-dimensional granular continuum. In particular, $Y_{2D}$ should not be confused with the Young's modulus obtained directly from a standard uniaxial test, as discussed, for example, in Eq.~(47) of ~\cite{misra2025emergence}.

At the macroscopic level, bimodulus effects are often summarized through a single compression-to-tension modulus ratio, commonly referred to in the literature as the bimodulus parameter \cite{wen2004analytical, parisi2015experimental}. In conventional bimodulus formulations, this parameter is classically written as
\begin{equation}
\bar{n}= \frac{E^{c}}{E^{t}},
\label{eq:nbar_def_beam}
\end{equation}
or, equivalently, through its inverse. This scalar ratio is widely used to quantify tension/compression asymmetry in bimodulus formulations and related structural analyses \cite{caporale2014micromechanical}.

In the present granular micromechanics framework, however, the distinction between tension and compression is introduced at the level of each pair interaction. More specifically, the state of an interaction depends on the sign of the normal displacement measure $u_\eta$, and therefore on the orientation $\theta$. With the adopted convention, $u_\eta>0$ identifies tensile interactions, whereas $u_\eta<0$ corresponds to compressive interactions. Accordingly, the normal and tangent stiffnesses are defined as
\begin{equation}
k_{\eta}=k_{\eta}^{t}\Theta(u_{\eta})+k_{\eta}^{c}\Theta(-u_{\eta}),
\qquad
k_{\tau}=k_{\tau}^{t}\Theta(u_{\eta})+k_{\tau}^{c}\Theta(-u_{\eta}),
\label{eq:asymmetric_stiffness_clean}
\end{equation}
where $\Theta(\cdot)$ is the Heaviside function, with $\Theta(x)=1$ for $x > 0$, $\Theta(x)=1/2$ for $x=0$ and $\Theta(x)=0$ for $x<0$. Here, $k_{\eta}^{t}$ and $k_{\tau}^{t}$ denote the tensile normal and tangent stiffnesses, while $k_{\eta}^{c}$ and $k_{\tau}^{c}$ denote their compressive counterparts. In the present numerical implementation, the switching rule in
Eq.~\eqref{eq:asymmetric_stiffness_clean} is employed through a piecewise
active-set procedure. The orientation integral is evaluated numerically using a finite set of interaction orientations. For a given strain field, each discrete
interaction orientation is classified according to the sign of
\(u_{\eta}\), and the corresponding values of \(k_{\eta}\) and
\(k_{\tau}\) are assigned. With this interaction-state distribution held
fixed, the stiffness tensor is evaluated from
Eq.~\eqref{eq:stiffness_tensor_rev}. The interaction states are subsequently
updated from the newly computed strain field and the procedure is repeated.
Within this discrete procedure, the orientation at which \(u_{\eta}=0\)
generally does not coincide exactly with any of the selected interaction
orientations.
Therefore, the switching orientation is not treated as a continuously moving
integration boundary when evaluating the constitutive response. Accordingly,
in the bimodulus case, Eq.~\eqref{eq:stiffness_tensor_rev} represents the
stiffness tensor associated with the current interaction-state distribution,
rather than the full derivative of
Eq.~\eqref{eq:energy_density_rev} including the derivative of the switching
orientation with respect to the macroscopic strain. The convention
\(\Theta(0)=1/2\) only specifies the value in the exceptional case in which
a discrete interaction lies exactly at the switching condition.

This orientation-dependent definition is important. A material point is generally not purely in tension or purely in compression in all directions: even under a macroscopic uniaxial tensile state, some interaction orientations may experience compression as a consequence of the transverse contraction induced by Poisson's effect. Therefore, the asymmetry of the granular model is more naturally described through the compression-to-tension ratios of the interaction stiffnesses,
\begin{equation}
n_{\eta}=\frac{k_{\eta}^{c}}{k_{\eta}^{t}},
\qquad
n_{\tau}=\frac{k_{\tau}^{c}}{k_{\tau}^{t}},
\qquad\Rightarrow\qquad
k_{\eta}^{t}=\frac{k_{\eta}^{c}}{n_{\eta}},
\quad
k_{\tau}^{t}=\frac{k_{\tau}^{c}}{n_{\tau}}.
\label{eq:two_ratios_clean}
\end{equation}
If idealized purely tensile or purely compressive states are considered, the effective parameters $Y_{2D}^{\alpha}$ and $\nu_{2D}^{\alpha}$, with $\alpha=t,c$, can be related to the corresponding interaction stiffnesses through Eq.~\eqref{eq:isotropic_stiffness_rev} as
\begin{equation}
\label{eq:isotropic_stiffness_tc}
k_\eta^{\alpha} = \frac{Y_{2D}^{\alpha}}{\pi L^2 (1 - \nu_{2D}^{\alpha})},
\qquad
k_\tau^{\alpha} = \frac{Y_{2D}^{\alpha}(1 - 3\nu_{2D}^{\alpha})}{4\pi L^2 (1 - \nu_{2D}^{\alpha})(1 + \nu_{2D}^{\alpha})},
\qquad \alpha=t,c.
\end{equation}
These effective quantities should not be interpreted as elastic constants measured directly from standard uniaxial tension or compression tests, but as macroscopic parameters associated with idealized uniform interaction states within the granular model.

The special case $n_{\eta}=n_{\tau}=\bar n$ recovers the single-parameter assumption adopted in \cite{nayeban2025analysis}. In this case, the second relation in Eq.~\eqref{eq:A_general} implies $\nu_{2D}^{t}=\nu_{2D}^{c}$.
It should be noted that, in the present first-gradient formulation, the
characteristic length \(L\) enters the macroscopic stiffness tensor
through the products \(L^2 k_{\eta}\) and \(L^2 k_{\tau}\) (see Eq.~\eqref{eq:stiffness_tensor_rev}). Consequently, the
experimental strain field does not identify \(L\) and the interaction stiffnesses
independently within the present first-gradient formulation. Instead, for a prescribed value of \(L\), the inverse procedure
identifies the corresponding interaction stiffnesses consistent with that
choice. If the characteristic length is rescaled as \(L \rightarrow \beta L\), with
\(\beta>0\), an equivalent first-gradient macroscopic response is obtained by
rescaling the interaction stiffnesses as
\(k_{\eta}\rightarrow k_{\eta}/\beta^{2}\) and
\(k_{\tau}\rightarrow k_{\tau}/\beta^{2}\).
Therefore, the stiffness values reported in this work should be interpreted as
conditional on the prescribed value of \(L\), while the stiffness ratios
\(n_{\eta}\) and \(n_{\tau}\) are unaffected by this common rescaling.

\subsection{Beam setting and neutral-axis description}\label{sec:beam_theory}

\begin{figure}[t]
  \centering
  \includegraphics[width=0.95\linewidth]{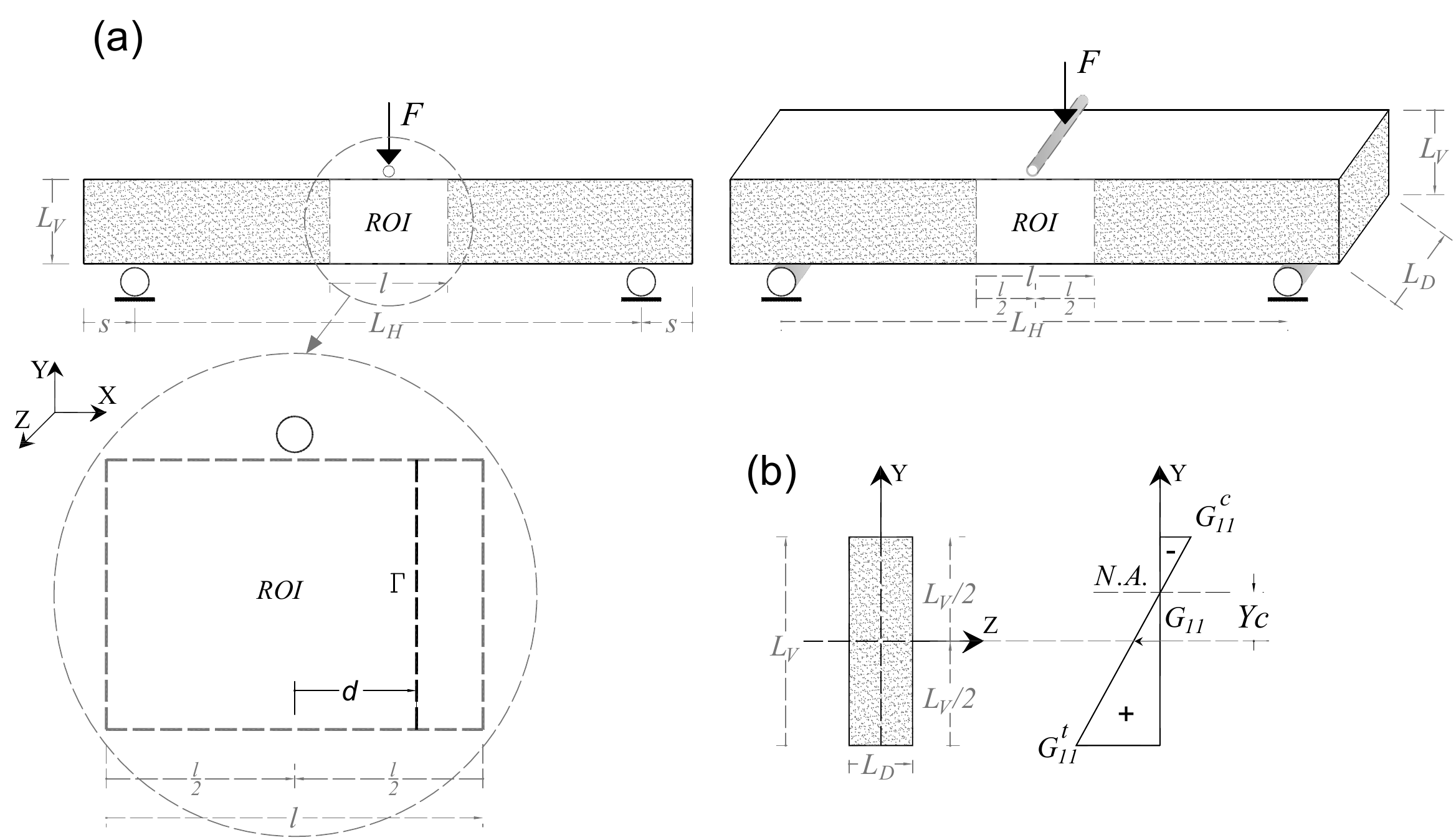}
\caption{Three-point bending setup and reference quantities. (a) Loading configuration and boundary conditions; $L_H$ denotes the clear span, $L_V$ the beam height, and $L_D$ the out-of-plane thickness. The selected DIC region of interest (ROI) and the vertical cross-section $\Gamma$ used for strain-profile comparisons are indicated; $d$ denotes the offset of $\Gamma$ from the loading point along the beam axis. (b) Schematic representation of the neutral-axis (N.A) position across the depth, together with the corresponding through-depth distributions of $G_{11}$. The quantity $Y_C$ denotes the deviation of the neutral axis from the section mid-depth.}
\label{fig:beam_axes}

\end{figure}

A prismatic beam subjected to three-point bending is considered in a two-dimensional setting.
This beam setting is consistent with the classical flexural framework in which Saint-Venant-type beam descriptions provide the natural reference for section-level stress and strain interpretation under bending \cite{barretta2014analogies, barretta2015analogies}. Related analogies between Kirchhoff-type plate formulations and Saint-Venant beam problems have also been explored in the literature for generalized and viscoelastic settings \cite{barretta2014exact}.
The reference frame is defined as follows (Figure~\ref{fig:beam_axes}): the $X$-axis is aligned
with the beam longitudinal axis, the $Y$-axis is the depth coordinate across the section,
and $Z$ denotes the out-of-plane thickness.
The components $G_{11}$, $G_{22}$, and $G_{12}$ denote, respectively, the longitudinal normal strain component in the $X$-direction, the transverse normal strain component in the $Y$-direction, and the in-plane shear strain component.

The neutral axis (N.A)  is defined as the locus in the cross-section along which the normal strain vanishes, i.e.,
$G_{11}=0$.
In the tension/compression symmetric case (classical linear-elastic behavior for a doubly-symmetric profile), the neutral axis coincides
with the mid-depth axis of the cross-section. For tension/compression asymmetric (bimodulus)
materials, the neutral axis shifts away from the mid-depth: the stiffer side (typically compression)
attracts the neutral axis, so that the tensile and compressive regions are no longer symmetric with respect
to the section centerline. This feature provides a convenient macroscopic signature of elastic
asymmetry and is exploited later in the identification procedure.

% Section-3 =================================================

\section{Experimental program and DIC analysis}\label{sec:DIC}

\subsection{Specimen, three-point bending setup, and loading protocol}\label{sec:exp_setup}

\begin{figure}[t]
    \centering
    \includegraphics[width=\textwidth]{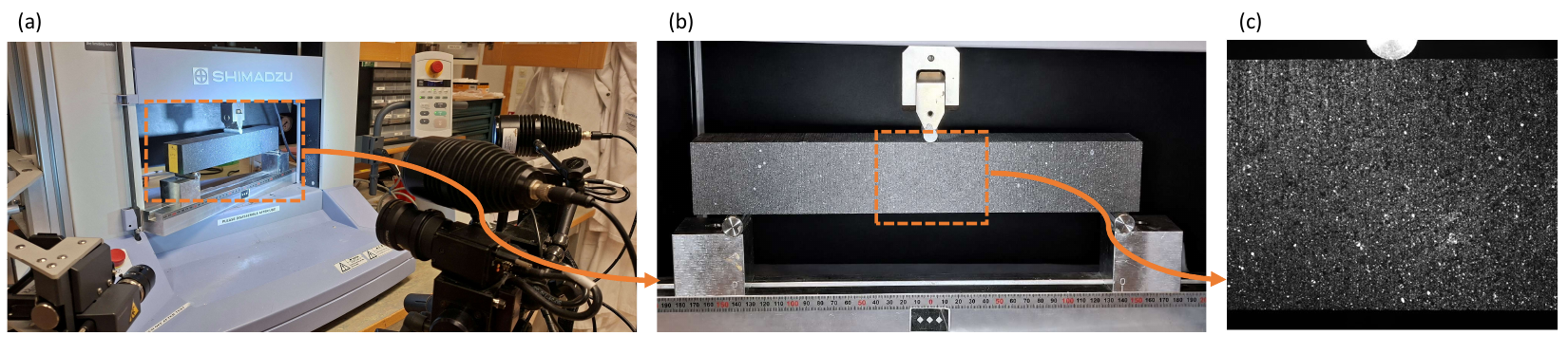}
    \caption{Experimental workflow of the three-point bending test and DIC acquisition: (a) overall view of the Shimadzu testing machine and DIC camera system; (b) close-up view of the beam specimen, supports, and loading nose; (c) representative camera image of the speckled specimen surface with the selected ROI. The dashed boxes and arrows indicate the progression from the global experimental setup to the local bending region and the ROI used in the DIC analysis.}
    \label{fig:exp_setup_combined}
\end{figure}

The experimental campaign is designed around simple three-point bending tests on prismatic beam specimens, with the aim of producing clear strain fields for subsequent inverse identification. Flexural testing is widely adopted as a practical experimental configuration for assessing the mechanical response of prismatic concrete and cement-based members under bending \cite{yildizdag2020three, rad2022externally}. Figure~\ref{fig:exp_setup_combined}(a) shows the actual laboratory setup, while Figure~\ref{fig:beam_axes}(a) provides the corresponding schematic layout, coordinate system, and geometric notation. The actual geometric dimensions of the beam specimen, corresponding to the notation introduced in Figure~\ref{fig:beam_axes}(a), are summarized in Table~\ref{tab:spec_geom}.

The tests were performed using a Shimadzu Autograph AGS-X testing machine equipped with a 10~kN load cell with $\pm$0.5\% precision. The specimen was simply supported at two locations offset by \(s\) from the specimen ends, as shown in Figure~\ref{fig:beam_axes}(a), and was loaded at midspan through a cylindrical loading nose. During the tests, the laboratory temperature and relative humidity were maintained at
$21.5 \pm 0.5\,^{\circ}\mathrm{C}$ and $40 \pm 2\%$, respectively. When required, repeated tests under controlled conditions were carried out to assess measurement repeatability and the robustness of the experimental procedure.

\subsection{DIC principle and experimental setup}\label{sec:DIC_setting}
Digital Image Correlation (DIC) is a non-contact optical technique that provides full-field
in-plane displacement and strain measurements by tracking the deformation of a speckle pattern
between digital images acquired during loading. In this work, DIC is used to obtain strain maps
in the $(X,Y)$ plane during three-point bending tests, with primary focus on the normal strain
component $G_{11}$.

The images were acquired with an Aramis DIC system equipped with 5 MP camera with fixed focus and a resolution of $2448 \times 2052$ pixels, leading to a strain accuracy of 0.01\%. The specimen reference frame follows Figure~\ref{fig:beam_axes}.
Since the identification relies on in-plane strain fields evaluated on a single observation plane, a two-dimensional DIC configuration was adopted. Accordingly, image acquisition was performed from a single camera view (single-side acquisition), as shown in Figure~\ref{fig:exp_setup_combined}(b), so that only the in-plane kinematics in the $(X,Y)$ plane is measured, while the out-of-plane component along $Z$ is not reconstructed.

In terms of post-processing, the in-plane strain fields were obtained independently from two
correlation tools, Ncorr (an open-source 2D DIC program implemented in MATLAB) and
PyDIC (a Python-based digital image correlation tool). The use of two workflows is motivated
by the need to cross-check the results and assess the robustness of the strain estimates. The adopted correlation parameters were selected following standard accuracy--resolution considerations. In particular, the subset size was chosen as a compromise between measurement noise and the ability to resolve localized strain gradients within the region of interest. The
strain-field outputs from both tools were then analyzed through a single custom MATLAB code
developed by the authors. This MATLAB pipeline imports the strain fields, applies a consistent
sampling and processing strategy, and generates the contour plots used throughout the paper.
In addition, the same pipeline provides the derived quantities required by the identification
procedure, including the extraction of the neutral-axis position from the \(G_{11}\) field.

\subsubsection{Region of interest and loading protocol}
\label{sec:DIC_roi}

A region of interest (ROI) is defined to restrict the correlation to a domain with stable image
quality and reliable texture. According to Figure~\ref{fig:beam_axes}(a), the ROI is selected at the center of the specimen span, with a width of $l$ as reported in Table~\ref{tab:spec_geom}, and
extending over the full beam height. The same ROI definition is adopted consistently for all tests
and for the comparison with numerical predictions.

Before the main loading stage, a small preload of \(50\,\mathrm{N}\) was applied
in order to ensure proper contact between the specimen, the supports, and the
cylindrical loading roller. As reported in the literature, the purpose of the
preload is to smooth out the possible unevenness of the contact surfaces,
thereby stabilizing the mechanical response at the initial stage of the test
\cite{bengtsson2024evaluating}. Following this initial conditioning, the
specimen is subjected to a load--unload--reload protocol. The load is first
increased up to \(3500\,\mathrm{N}\), then reduced back to the zero-load level,
and subsequently reapplied up to \(3500\,\mathrm{N}\) in a second loading phase.
The peak load of \(3500\,\mathrm{N}\) was selected in such a way that the maximum
strain response within the region of interest remained compatible with an
essentially elastic interpretation of the test, thus limiting the influence of
possible non-elastic effects on the identification procedure. This procedure
allows for the assessment of the reproducibility of the measured response and
the consistency of the strain fields obtained from DIC. For the inverse
identification, the comparison between experimental and numerical strain fields
is performed between the zero-load configuration and the peak-load configuration
of the second loading phase. No explicit rigid-body rotation correction is
applied in the present post-processing.

The tests were conducted under displacement-controlled conditions with a reference loading rate of 2 mm/min.  In order to assess the influence of possible rate-dependent (viscous) effects, a series of tests was performed at different loading rates. The loading speed was progressively reduced down to values two orders of magnitude smaller than the reference rate, and the resulting responses were compared to verify convergence toward a rate-independent behavior.
The comparison of the resulting strain fields revealed negligible variations, indicating that the response is essentially rate-independent within the investigated range.

\subsubsection{DIC outputs and neutral-axis extraction}\label{sec:DIC_outputs}
The DIC analysis provides the in-plane strain components. Among these quantities,
the present study focuses on the normal strain component $G_{11}$, which represents the
strain along the beam $X$ axis. Although \(G_{12}\) and \(G_{22}\) can also be obtained from the DIC analysis,
their magnitudes are significantly smaller than that of \(G_{11}\), resulting
in lower signal-to-noise ratios and a greater influence of experimental noise. As a result, they are much more sensitive to measurement noise and do not
provide a sufficiently accurate and precise basis for the present identification procedure.
For this reason, the inverse analysis is built primarily on the comparison of the $G_{11}$
field.

The neutral axis is extracted directly from the measured $G_{11}$ field within the
selected ROI, indicated in Figure~\ref{fig:beam_axes}. Since the neutral axis is defined as
the locus across the beam depth where the normal strain changes sign, its local position is
identified from the sign change of $G_{11}$ along the depth direction. More specifically,
for each position $X$ within the ROI, the location at which $G_{11}$ changes sign is
interpreted as a local neutral-axis coordinate, denoted by $Y_{NA}(X)$.
By repeating this operation over the full ROI, a set of local neutral-axis positions
$Y_{NA}(X)$ is obtained. The local neutral-axis coordinates $Y_{NA}(X)$ obtained within the ROI were summarized by their
average value $\overline{Y}_{NA}$. The neutral-axis deviation $Y_C$ was then defined as the offset
of $\overline{Y}_{NA}$ with respect to the section mid-depth, consistently with the convention of
Figure~\ref{fig:beam_axes}(b). This quantity provides a compact
experimental measure of the departure of the neutral axis from the beam mid-depth and is
later used in the initial stage of the identification procedure.

\subsection{No-load assessment of DIC accuracy and precision}
\label{sec:DIC_noload}

Before the mechanical tests, the DIC measurement quality was assessed under a no-load condition to obtain a reference indicator of the measurement stability associated with the optical setup and the adopted correlation/post-processing settings. After calibration and once the test scene was fixed (camera and specimen position, focus, and lighting), a sequence of five images of the undeformed specimen was acquired while keeping the applied load equal to zero. One image was retained as the reference, while the remaining four images were correlated against it, resulting in N=4 no-load correlations. The same acquisition parameters later used during the bending tests were maintained.

Since no physical deformation is expected in this condition, any nonzero
measured displacement or strain is associated with measurement uncertainty
arising from image noise, speckle quality, lighting fluctuations, and correlation
settings. Following common practice in DIC validation \cite{sadek2005closure}, accuracy
is interpreted here as the systematic bias of the measured strain field under
zero deformation, while precision is associated with the repeatability of the
measurements across repeated no-load acquisitions. In the present work, these
quantities are evaluated as global indicators based on the ROI-averaged
\(G_{11}\) field. Therefore, they should not be interpreted as pointwise estimates of the local
full-field DIC noise entering the normalized root-mean-square error (NRMSE)
objective function \(J\) used in the identification procedure.

The no-load measurements were taken in the same region of interest (ROI)
later used for the identification procedure. For each no-load image
correlation \(k=1,\ldots,N\), the strain components \(G_{ij}\) were evaluated
over the ROI. In the present study, particular attention is given to the
normal strain component \(G_{11}\). Let \(G_{11}^{(k)}(X)\) denote the measured
\(G_{11}\) field for the \(k\)-th no-load correlation at a point \(X\) in the
ROI. Here, \(k=1,\ldots,N\) denotes the no-load correlation index, while
\(a=1,\ldots,m\) denotes the DIC point index within the ROI. The average
\(G_{11}\) value for each no-load correlation is then computed as

\begin{equation}
\overline{G}_{11}^{(k)}
=
\frac{1}{m}
\sum_{a=1}^{m}
G_{11}^{(k)}(a).
\label{eq:noload_avg}
\end{equation}
Here, \(k=1,\ldots,N\) denotes the no-load correlation index, with
\(N=4\), and \(a=1,\ldots,m\) denotes the DIC point index, with
\(m=9585\).
and then evaluate the no-load bias and scatter over the full sequence as
\begin{equation}
\mu_{G_{11}}
=
\frac{1}{N}
\sum_{k=1}^{N}\overline{G}_{11}^{(k)},
\qquad
\sigma_{G_{11}}
=
\sqrt{
\frac{1}{N-1}
\sum_{k=1}^{N}
\left(
\overline{G}_{11}^{(k)}-\mu_{G_{11}}
\right)^2
}.
\label{eq:noload_acc_prec}
\end{equation}
Here, \(\mu_{G_{11}}\) represents the bias of the ROI-averaged no-load
\(G_{11}\) field, while \(\sigma_{G_{11}}\) quantifies the scatter of this
ROI-averaged quantity across the repeated no-load correlations. The correlations
are evaluated with respect to the same reference image and are therefore used as
repeated no-load correlations, rather than as fully independent measurements.
The same procedure can be applied to other strain components if necessary.

The no-load indicators computed from Eqs.~\eqref{eq:noload_avg} and
\eqref{eq:noload_acc_prec} for the normal strain component \(G_{11}\) are
summarized in Table~\ref{tab:noload_dic}. These values are later used as a global reference for the experimental
uncertainty and parameter-resolution analysis of the inverse identification.

\subsection{Material screening and final selection}
\label{sec:material_selection}

A preliminary screening was performed to identify a material suitable for the inverse identification of the granular model parameters from DIC strain fields in three-point bending.

Several candidate materials previously considered in the literature for bimodulus experimental testing
\cite{misseri2022rammed, pastor2019elastic, zerdzicki2025compressive, yao2016elastic, key-1}
were reviewed and preliminarily evaluated against three operational criteria: the material had to develop a measurable elastic strain field under three-point bending, remain within the elastic regime throughout the DIC acquisition window, and exhibit a sufficient tension/compression stiffness contrast to make the inverse identification meaningful.

The screened materials included gypsum/plaster, 3D-printed polymers such as PLA, cast iron, natural building stones such as marble, and lime mortar (NHL~3.5). These materials were not retained for different reasons. Gypsum and plaster were considered too brittle and sensitive to specimen preparation, leading to limited repeatability before cracking. 3D-printed polymers were affected by printing direction, infill pattern, and possible rate-dependent\cite{afshar2025predicting} and nonlinear effects\cite{afshar2023nonlinear}, which would introduce additional anisotropy and uncertainty. Natural building stones were affected by material variability and by the limited availability of reliable tensile/compressive elastic data for the specific specimens. Cast iron was not adopted because of its brittle tensile response and limited measurable elastic strain range before damage initiation. NHL~3.5 lime mortar was initially considered because of its documented tension/compression contrast; however, preliminary numerical checks indicated that tensile cracking could occur before a sufficiently extended measurable elastic strain field developed within the selected ROI.

For each candidate, the reported elastic properties in tension and compression were translated into the corresponding granular stiffness parameters through the constitutive relations introduced in Section~\ref{sec:Methodology}. These parameters were then used as inputs to forward numerical simulations of the target
three-point bending configuration in order to estimate the expected strain levels and to check
the elastic admissibility of the response. The numerical outputs were checked against the
screening criteria, with particular attention to the magnitude of $G_{11}$ within the ROI and
to the expected onset of cracking or non-elastic effects under the considered bending
configuration.

Based on the screening outcomes, a commercial double-impregnated graphite
grade, EG-92E, was selected as the reference material for the present
identification study. The nominal properties of this graphite grade are
summarized in Table~\ref{tab:graphite_props}. These values were used to
define the material scale and to construct the initial stiffness estimate. A tensile Young's modulus was not reported in the supplier data and was
therefore not used as an independent initialization input. Instead, the
initial tensile interaction stiffnesses were obtained from the
neutral-axis-based asymmetry ratio.
The selected material provides strain levels within the ROI that are suitable
for DIC acquisition, remains within the elastic regime over the loading range
considered, and presents a sufficient tension/compression stiffness contrast
to make the inverse identification meaningful. In the identification procedure,
the Poisson's ratio is assigned at the initialization stage; a more accurate
determination may be addressed in a dedicated identification procedure.

The geometry of the specimen and the adopted DIC region of interest are summarized in
Table~\ref{tab:spec_geom}, consistent with the notation introduced in
Figure~\ref{fig:beam_axes}(a). In particular, $L_H$, $L_V$, and $L_D$ denote the specimen clear span, height, and thickness, respectively, while $s$ denotes the support offset.

\begin{table}[t]
\centering
\caption{Nominal properties of the selected graphite grade EG-92E.}
\label{tab:graphite_props}
\begin{tabular}{llc}
\toprule
\textbf{Symbol} & \textbf{Property} & \textbf{Value} \\
\midrule
-- & Grade & EG-92E, double impregnated graphite \\
$\rho_b$ & Bulk density & $1.75~\mathrm{g/cm^3}$ \\
$d_g$ & Maximum grain size & $0.8~\mathrm{mm}$ \\
$f_f$ & Flexural strength, along grain & $24~\mathrm{MPa}$ \\
$f^c$ & Compressive strength, along grain & $40~\mathrm{MPa}$ \\
$f^t$ & Tensile strength, along grain & $17~\mathrm{MPa}$ \\
$E^{c}$ & Compressive Young's modulus, along grain & $14~\mathrm{GPa}$ \\
\bottomrule
\end{tabular}
\end{table}

\begin{table}[t]
\centering
\caption{Geometric parameters of the specimen and the adopted DIC region of interest. The notation is consistent with Figure~\ref{fig:beam_axes}(a). All dimensions are given in mm.}
\label{tab:spec_geom}
\begin{tabular}{llc}
\toprule
\textbf{Symbol} & \textbf{Description} & \textbf{Value [mm]} \\
\midrule
$L_V$ & Specimen height (along $Y$) & 52 \\
$L_H$ & Clear span (along $X$) & 280 \\
$L_D$ & Out-of-plane thickness (along $Z$) & 43 \\
$s$   & Support offset (distance from support to beam end) & 15 \\
$l$   & ROI width (along $X$) & 80 \\
$d$   & Profile distance from the loading point & 14 \\
\bottomrule
\end{tabular}
\end{table}

\section{Identification procedure and results}\label{sec:results}

\subsection{Identification workflow}
The identification strategy combines the experimentally measured normal strain ($G_{11}$) field with the granular micromechanics-based numerical model in a step-by-step procedure. The identification problem is treated as a nonlinear regression problem. In this setting, different parameter sets may be mathematically admissible from the point of view of the objective function, although not all of them are mechanically meaningful. Therefore, a physically informed initialization is introduced before the iterative optimization stage. The local neutral-axis coordinates extracted from the DIC strain field are first summarized through their arithmetic mean, $\overline{Y}_{\mathrm{NA}}$, which provides a stable macroscopic measure of the neutral-axis position. The corresponding neutral-axis shift is then used as a physically informed estimate of the tension/compression asymmetry in bending and provides the basis for constructing a realistic initial estimate of the bimodulus parameters.
 This initialization helps guide the optimization toward mechanically admissible regions of the parameter space.

Based on this DIC-informed estimate and on the nominal material properties, an initial set of stiffness parameters is constructed and assigned to the numerical model. Starting from this initial parameter set, the identification is then performed through a direct full-field comparison between the numerical and experimental $G_{11}$ distributions over the selected region of interest. The discrepancy
is quantified by the objective function \(J\), defined as the normalized
root-mean-square error (NRMSE).
The optimization is carried out by means of a hybrid strategy. In the first stage, a pattern-search procedure is used to explore the parameter space without requiring derivatives of the objective function and to identify a suitable neighborhood of the optimum. In the second stage, the parameter set obtained from the pattern-search stage is further refined by a quasi-Newton local procedure, aimed at improving convergence in the vicinity of the minimum. In this way, the neutral-axis shift provides the physically informed starting point, whereas the final identification is governed by the minimization of the full-field discrepancy between the numerical and experimental $G_{11}$ fields.

\subsection{Experimental strain field and initial parameter set}

\begin{figure}[t]
\centering
\includegraphics[width=0.97\textwidth]{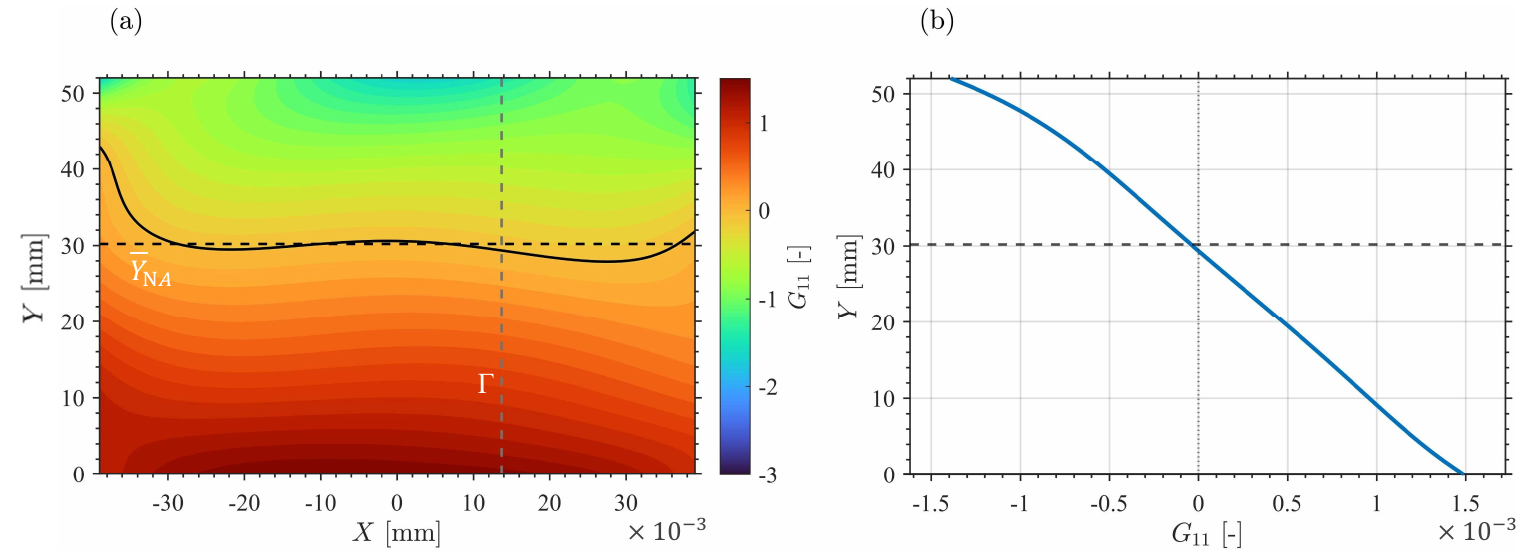}
\caption{
(a) Full-field distribution of the normal strain component $G_{11}$ over the selected Region Of Interest (ROI). 
The selected ROI corresponds to the DIC region defined in
Figure~\ref{fig:beam_axes}(a) and Table~\ref{tab:spec_geom}, and extends over
the full specimen height.
The dashed horizontal line denotes the average neutral-axis position $\overline{Y}_{\mathrm{NA}}$, while the solid curve represents the local neutral-axis coordinates $Y_{\mathrm{NA}}(X)$ extracted along the beam axis. The vertical dashed line indicates the location of the profile $\Gamma$ used for one-dimensional evaluation.
(b) Through-depth distribution of $G_{11}$ along the selected vertical profile $\Gamma$. The zero-crossing point provides a local estimate of the neutral-axis position.}
\label{fig:G11_field_profile}
\end{figure}

The experimental normal strain field $G_{11}$ obtained from DIC over the selected region of interest (ROI) is shown in Figure~\ref{fig:G11_field_profile}(a). The corresponding through-depth distribution along the vertical profile $\Gamma$ is reported in Figure~\ref{fig:G11_field_profile}(b).  The profile $\Gamma$ is intentionally selected at a finite distance from the loading point, rather than directly beneath the loading nose, in order to avoid the region most strongly affected by localized stress-transfer effects associated with the concentrated load. In this way, the comparison focuses on the bending-dominated response, which is more representative of the assumptions underlying the present first-gradient continuum model.
From the measured strain field, the neutral axis is identified within the ROI as the locus
along which the normal strain changes sign. For each position $X$ along the beam axis, a local neutral-axis coordinate
$Y_{\mathrm{NA}}(X)$ is extracted. The resulting set of local neutral-axis coordinates is
then summarized by its average value,
\begin{equation}
\overline{Y}_{\mathrm{NA}}
=
\frac{1}{N_X}
\sum_{r=1}^{N_X}
Y_{\mathrm{NA}}(X_r),
\label{eq:average_neutral_axis}
\end{equation}
where $X_r$, with $r=1,\ldots,N_X$, denotes the sampling positions along the beam axis
within the ROI. The quantity $\overline{Y}_{\mathrm{NA}}$ provides a global estimate of the
neutral-axis position for the considered configuration. The neutral-axis deviation \(Y_C\) is subsequently defined as the signed offset of
\(\overline{Y}_{\mathrm{NA}}\) from the section mid-depth, using a sign convention
opposite to the positive \(Y\)-direction shown in
Figure~\ref{fig:beam_axes}(b). Accordingly, \(Y_C\) is negative for an upward
shift of the neutral axis and positive for a downward shift.

Following the classical bimodulus beam formulation
\cite{key-1,parisi2015experimental}, the neutral-axis deviation can be related
to the macroscopic bimodulus ratio \(\bar{n}\) through
\begin{equation}
Y_C=\frac{L_V\left(1-\sqrt{\bar{n}}\right)}
{2\left(1+\sqrt{\bar{n}}\right)}.
\label{eq:YC_nbar_beam}
\end{equation}
This relation is used here only to provide an initial estimate of \(\bar{n}\)
from the experimentally measured neutral-axis position.

The measured value of \(Y_C\) is used to obtain an initial estimate of the
macroscopic bimodulus ratio \(\bar{n}\) through
Eq.~\eqref{eq:YC_nbar_beam}. This physically informed estimate provides a
realistic starting point for the nonlinear regression procedure and helps guide
the subsequent optimization toward mechanically meaningful parameter values.
An initial set of stiffness parameters is then constructed from the
supplier-reported nominal compressive Young's modulus listed in
Table~\ref{tab:graphite_props}, together with the Poisson's ratio assigned at
the initialization stage. In the present calculations, the characteristic
length used in the first-gradient model and the Poisson's ratio adopted at the
initialization stage were prescribed as \(L=10~\mathrm{mm}\) and
\(\nu_{2D}=0.16\), respectively.
This modulus is not interpreted as the final effective modulus of the granular
continuum. As discussed in Section~\ref{sec:Methodology}, \(Y_{2D}\) is an
effective two-dimensional parameter and should not be confused with a Young's
modulus measured directly from standard uniaxial test. Here, the
supplier-reported value is used only to define the initial compressive
stiffness scale, because the compressive branch is the dominant reference for
the compression side of the bending response.
For this purpose, the supplier-reported compressive modulus \(E^{c}\) is
converted into the corresponding two-dimensional quantity by multiplying it by
the out-of-plane specimen thickness \(L_D\), i.e.,
\(Y_{2D}^{c}=E^{c}L_D\). This conversion provides only an initial estimate.
Therefore, the stiffnesses estimated from \(Y_{2D}^{c}\) should be regarded
only as initial values and are subsequently corrected by the full-field
optimization.
The corresponding initial compressive stiffnesses \(k_{\eta}^{c}\) and
\(k_{\tau}^{c}\) can be obtained from the relations introduced in
Section~\ref{sec:Methodology} (see Eq.~\eqref{eq:isotropic_stiffness_tc}).
For the considered specimen, \(L_V=52~\mathrm{mm}\), while the experimentally
measured average neutral-axis position is approximately
\(\overline{Y}_{\mathrm{NA}}=30.20~\mathrm{mm}\). According to the adopted sign
convention, this corresponds to a neutral-axis deviation
\(Y_C=L_V/2-\overline{Y}_{\mathrm{NA}}\approx-4.20~\mathrm{mm}\), which gives
\(\bar{n}=1.92\) through Eq.~\eqref{eq:YC_nbar_beam}. For the initial parameter
set ,\(\mathbf{q}^{0}\), reported in the  Table~\ref{tab:identification_steps}, this
asymmetry ratio is assigned equally to both interaction channels, i.e.,
\(n_{\eta}=n_{\tau}=\bar{n}=1.92\). The tensile stiffnesses are then obtained
from Eq.~\eqref{eq:two_ratios_clean}. This procedure defines the initial
parameter set
\((k_{\eta}^{c},k_{\eta}^{t},k_{\tau}^{c},k_{\tau}^{t})\).

The corresponding numerical solution is obtained by solving the model under
the same geometry, boundary configuration, and final load level used for the
DIC comparison. The predicted
$G_{11}$ field is then evaluated over the same ROI and along the same vertical profile
$\Gamma$ used for the experimental measurements. The resulting value of \(\bar{n}\), together with the associated initial
stiffness parameters defining \(\mathbf{q}^{0}\), is summarized in
Table~\ref{tab:identification_steps} and serves as the starting point for
the subsequent identification procedure.
In the present first-gradient identification, the characteristic length
\(L\) was kept fixed and was not treated as an additional optimization
variable.
Accordingly, the optimization was restricted to four independent quantities:
the compressive normal and tangent stiffnesses,
\(k_{\eta}^{c}\) and \(k_{\tau}^{c}\), and the two compression-to-tension stiffness
ratios, \(n_{\eta}\) and \(n_{\tau}\). The reported stiffnesses are therefore
the values associated with the prescribed length scale \(L\). This choice is specific to the first-gradient setting considered here. In a
strain-gradient extension, the characteristic length would contribute to
higher-order terms and would no longer act only as a scaling factor of the
first-gradient stiffness tensor. In that case, \(L\) should be identified as an
additional parameter together with the interaction stiffnesses.

\begin{table}[htb]
\centering
\small
\setlength{\tabcolsep}{1pt}
\caption{Initial and final physical parameter sets used in the identification
procedure. Here, \(\mathbf{q}^{0}\) denotes the initial parameter set, while
\(\mathbf{q}^{\mathrm{opt}}\) denotes the final identified parameter set.}
\label{tab:identification_steps}
\begin{tabular}{@{}l@{\hspace{6pt}}ccc|cccc}
\toprule
& \multicolumn{3}{c|}{Bimodulus parameters (–)}
& \multicolumn{4}{c}{Stiffness parameters ($\mathrm{N/m^3}$)} \\
\cmidrule(r){2-4} \cmidrule(l){5-8}
& $\bar{n}$ & $n_{\eta}$ & $n_{\tau}$
& $k_{\eta}^{c}$ & $k_{\eta}^{t}$ & $k_{\tau}^{c}$ & $k_{\tau}^{t}$ \\
\midrule
$\mathbf{q}^{0}$
& 1.92
& 1.92
& 1.92
& $2.28\times10^{12}$
& $1.19\times10^{12}$
& $2.56\times10^{11}$
& $1.33\times10^{11}$ \\

$\mathbf{q}^{\mathrm{opt}}$
& -
& 1.60
& 1.88
& $1.86\times10^{12}$
& $1.16\times10^{12}$
& $1.56\times10^{11}$
& $8.28\times10^{10}$ \\
\bottomrule
\end{tabular}
\end{table}

\subsubsection{No-load DIC accuracy and precision}

The no-load assessment introduced in Section~\ref{sec:DIC_noload} was based on
five unloaded images. One image was retained as the reference, while the remaining
four images were correlated against it, resulting in $N=4$ no-load correlations.
For each correlation, the normal strain component $G_{11}$ was extracted over the ROI
and its spatial average $\overline{G}_{11}^{(k)}$ was computed according to
Eq.~\eqref{eq:noload_avg}. The no-load bias and scatter were then evaluated using
Eq.~\eqref{eq:noload_acc_prec}. The resulting values are summarized in
Table~\ref{tab:noload_dic}.

\begin{table}[t]
\centering
\caption{No-load DIC indicators for the normal strain component $G_{11}$.
The spatial averages $\overline{G}_{11}^{(k)}$ are computed according to
Eq.~\eqref{eq:noload_avg}, while $\mu_{G_{11}}$ and $\sigma_{G_{11}}$ are
evaluated according to Eq.~\eqref{eq:noload_acc_prec}.}
\label{tab:noload_dic}

\renewcommand{\arraystretch}{1.15}
\setlength{\tabcolsep}{10pt}
\begin{tabular}{cccccc}
\hline
Quantity & Value
& Quantity & Value
& Quantity & Value \\
\hline
$\overline{G}_{11}^{(1)}$
& $-2.54 \times 10^{-5}$
& $\overline{G}_{11}^{(2)}$
& $-3.66 \times 10^{-5}$
& $\overline{G}_{11}^{(3)}$
& $-1.17 \times 10^{-5}$ \\

$\overline{G}_{11}^{(4)}$
& $-2.13 \times 10^{-5}$
& $\mu_{G_{11}}$
& $-2.37 \times 10^{-5}$
& $\sigma_{G_{11}}$
& $\phantom{-}1.03 \times 10^{-5}$ \\
\hline
\end{tabular}
\end{table}
The results show that the measured no-load normal strain remains within the
order of \(10^{-5}\). Therefore, the adopted DIC configuration provides a stable
global strain estimate in the absence of mechanical loading. In the following identification results, the values reported in Table~\ref{tab:noload_dic} are used only as a global reference level for interpreting the magnitude of the
measured strain field.

\subsection{Optimization procedure and parameter evolution}

\subsubsection{Numerical implementation}

The forward solver employed in the present study follows the first-gradient granular micromechanics formulation summarized in Section~\ref{sec:Methodology} and introduced in detail in ~\cite{nayeban2025analysis}. For a given trial parameter set, the unknown field is the displacement field of the three-point bending specimen under the prescribed boundary and loading conditions. For a fixed interaction-state distribution, the finite-element problem is
solved in COMSOL from the stationarity of the corresponding total potential
energy, whose elastic contribution is obtained from the granular
strain-energy density \(U\). The interaction-state distribution is then
updated externally through the MATLAB--COMSOL iterative procedure described
below. The local energy density is defined in Eq.~\eqref{eq:energy_density_rev}, where the normal and tangent displacements are computed from the Green--Saint-Venant strain components according to Eq.~\eqref{eq:micro_measures_rev}. The strain tensor itself is obtained from the deformation gradient through Eq.~\eqref{eq:F_Gij_def_rev}.
After substitution of the displacement measures into the energy density, the macroscopic elastic energy is written in the compact quadratic form of Eq.~\eqref{eq:compact_energy_rev}, with the fourth-order stiffness tensor defined by Eq.~\eqref{eq:stiffness_tensor_rev}. In the bimodulus case, this stiffness tensor is not fixed a priori, because the interaction stiffnesses $k_\eta$ and $k_\tau$ depend on the sign of the normal displacement $u_\eta$, as prescribed in Eq.~\eqref{eq:asymmetric_stiffness_clean}. Since $u_\eta$ depends on the strain field, and the strain field depends on the displacement solution, the constitutive response must be updated during the numerical solution.
The numerical implementation is therefore based on a coupled MATLAB--COMSOL workflow. COMSOL solves the finite-element boundary-value problem for the current set of stiffness parameters and returns the displacement and strain fields. MATLAB then evaluates $u_\eta$ at each material point and for each interaction orientation, assigns the tensile or compressive values of $k_\eta$ and $k_\tau$ according to Eq.~\eqref{eq:asymmetric_stiffness_clean}, and reconstructs the corresponding stiffness tensor through Eq.~\eqref{eq:stiffness_tensor_rev}. This updated stiffness tensor is then passed back to COMSOL for the next numerical solution.

The experimental three-point bending configuration was reproduced in the
numerical model by applying the loading and support conditions over finite
boundary segments rather than at mathematical points. The vertical load at
midspan was prescribed as a uniformly distributed boundary load over a segment
of width $1~\mathrm{mm}$, and the supports were represented over boundary
segments of the same width, in order to better approximate the finite contact
regions associated with the experimental loading nose and supports. The load
was applied incrementally in five steps of $700~\mathrm{N}$, reaching a maximum
value of $3500~\mathrm{N}$. The incremental loading procedure is required because the interaction
stiffnesses, and consequently the stiffness tensor, depend on the evolving
strain field through the sign of the normal displacement
\(u_{\eta}\). Therefore, the constitutive state is updated sequentially as
the load is increased, following the computational strategy described in
detail in \cite{nayeban2026numerical}. The number of load increments was
selected based on a convergence analysis in which the evolution of the
stiffness tensor was monitored as the number of increments was progressively
increased. Since only negligible variations were observed when five or more
load increments were used, five increments were adopted as a suitable
compromise between numerical accuracy and computational cost. At the beginning of the computation, the local
tension/compression state of the interactions is not yet known; therefore, an
initial symmetric estimate of the interaction stiffnesses is used to obtain a
first non-trivial strain field. This strain field is exported to MATLAB, where
the interaction stiffnesses are updated according to the sign of $u_\eta$, and
the finite-element problem is then solved again. The same update procedure is
repeated over the loading increments until the prescribed final loading level
is reached. The resulting numerical $G_{11}$ field over the selected ROI is
then extracted and used in the inverse identification loop, while the field at
the target load level is compared with the DIC measurements.

For each trial parameter set, the forward model generates a numerical
\(G_{11}\) field over the selected ROI. This field is directly compared with
the corresponding experimental \(G_{11}\) field obtained from DIC. Let
\(a=1,\ldots,m\) denote the sampling points within the ROI, where \(m\) is the
total number of points used in the comparison. The quantity
\(G_{11}^{\mathrm{obs}}(a)\) denotes the experimentally observed normal strain
at point \(a\), whereas \(G_{11}^{\mathrm{mod}}(a;\mathbf{q})\) denotes the
numerical normal strain predicted by the forward model at the same point for a
given trial parameter set $
\mathbf{q}$. The discrepancy between the numerical
and experimental fields is quantified using the normalized root-mean-square
error (NRMSE), which is adopted as the objective function and is hereafter
denoted by \(J\). Thus,
\begin{equation}
J(\mathbf{q})
\equiv
\mathrm{NRMSE}(\mathbf{q})
=
\frac{
\sqrt{\dfrac{1}{m}\sum_{a=1}^{m}
\left(G_{11}^{\mathrm{mod}}(a;\mathbf{q})
-G_{11}^{\mathrm{obs}}(a)\right)^2}
}{
\sqrt{\dfrac{1}{m}\sum_{a=1}^{m}
\left(G_{11}^{\mathrm{obs}}(a)\right)^2}
}
\times 100, \qquad\mathbf{q}
=
\begin{bmatrix}
k_{\eta}^{c} &
k_{\eta}^{t} &
k_{\tau}^{c} &
k_{\tau}^{t}
\end{bmatrix}^{T} .
\label{eq:nrmse}
\end{equation}
In this form, the difference between the numerical and experimental strain
fields is normalized by the root-mean-square value of the observed
\(G_{11}\) field, so that \(J\) is expressed as a percentage of the
experimental strain level. At this stage, the comparison is performed
directly on the full-field strain distribution, rather than on the
neutral-axis position. The objective function \(J\) is therefore used to
guide the calibration of the model parameters.
\subsubsection{Hybrid optimization procedure}

\begin{figure}[t]
    \centering
    \includegraphics[width=\textwidth]{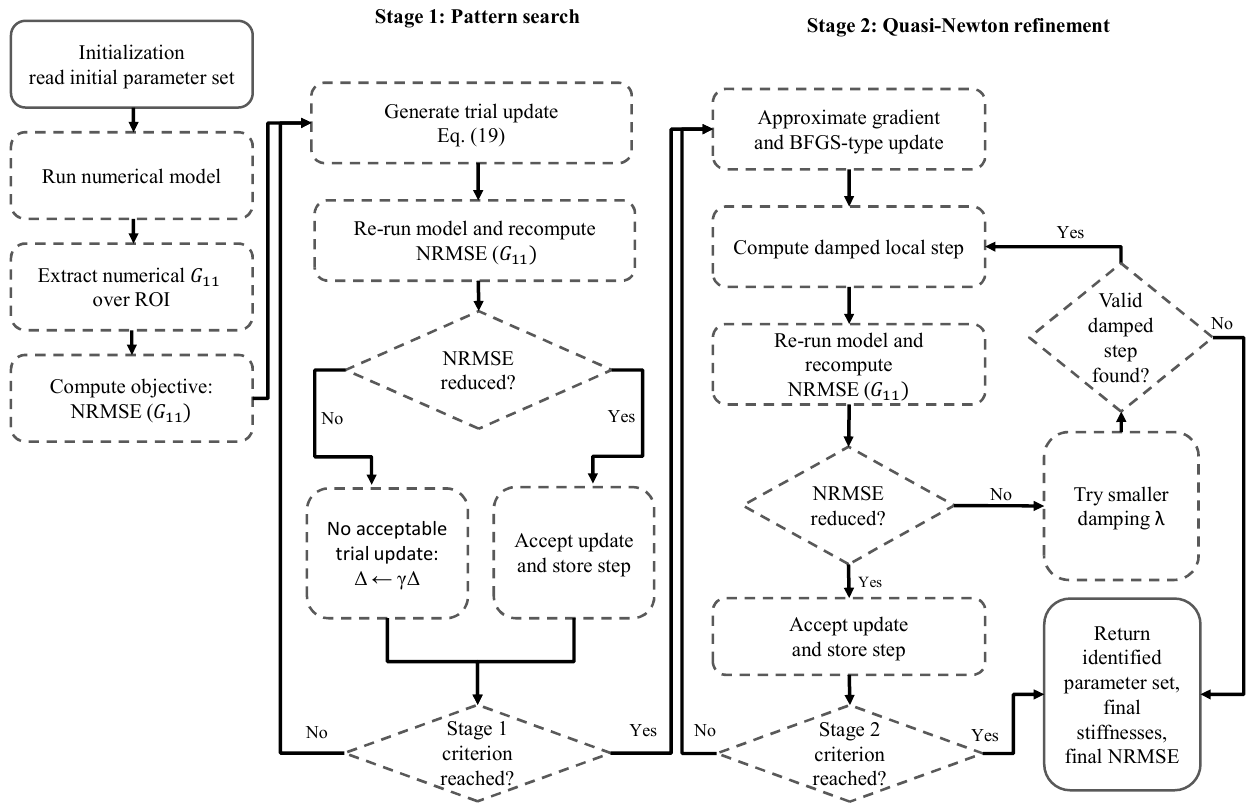}
   \caption{Flowchart of the hybrid optimization procedure adopted for the
identification process. Stage~1 consists of a pattern-search exploration based
on the objective function \(J\), defined as the normalized root-mean-square
error (NRMSE) of the full-field \(G_{11}\) comparison, whereas Stage~2 performs
a quasi-Newton refinement starting from the best parameter set obtained in the
first stage.}
    \label{fig:optimization_flowchart}
\end{figure}

Starting from the initial estimate described above, the identification is
performed by optimizing the compressive normal and tangent stiffnesses
together with the corresponding tension/compression asymmetry ratios. For improved numerical conditioning, the stiffness parameters are
nondimensionalized using fixed reference values. The optimization vector is
defined as
\begin{equation}
\mathbf{p}
=
\begin{bmatrix}
\widetilde{k}_{\eta}^{c} &
\widetilde{k}_{\tau}^{c} &
n_{\eta} &
n_{\tau}
\end{bmatrix}^{T}.
\qquad
\widetilde{k}_{\eta}^{c}
=
\frac{k_{\eta}^{c}}
{10^{12}\,\mathrm{N/m^3}},
\qquad
\widetilde{k}_{\tau}^{c}
=
\frac{k_{\tau}^{c}}
{10^{11}\,\mathrm{N/m^3}}.
\label{eq:opt_parameters}
\end{equation}
Thus, all four components of \(\mathbf{p}\) are dimensionless.
Here, \(k_{\eta}^{c}\) and \(k_{\tau}^{c}\) denote the compressive
normal and tangent stiffnesses, respectively, while \(n_{\eta}\) and
\(n_{\tau}\) are the corresponding compression-to-tension asymmetry ratios.
At each trial state, the physical stiffnesses are reconstructed from the
scaled variables and the tensile stiffnesses are then obtained according to
the constitutive relations introduced in Section~\ref{sec:Methodology}
(see Eq.~\eqref{eq:two_ratios_clean}).

The optimization is implemented in an automatic MATLAB-based workflow. For
each trial parameter set, the numerical model is re-run, the corresponding
\(G_{11}\) field is extracted over the selected ROI, and \(J\) defined in
Eq.~\eqref{eq:nrmse} is recomputed by comparison with the experimental field.
The resulting \(J\) is then used to accept or reject the trial update. A
schematic representation of the complete workflow is provided in
Figure~\ref{fig:optimization_flowchart}.

The optimization is carried out in two stages. In the first stage, a
pattern-search procedure is used to explore the parameter space without
computing derivatives of the objective function. This choice is suitable for
the present MATLAB--COMSOL workflow, since each trial parameter set requires
a complete numerical simulation and the objective function is evaluated from the resulting full-field
\(G_{11}\) distribution.
At the beginning of the \(k\)-th outer pattern-search iteration, a current
parameter vector \(\mathbf{p}_{\mathrm{cur}}^{(k)}\) and a step-size vector
\(\bm{\Delta}^{(k)}\) are available. The parameters are explored sequentially.
For the \(j\)-th parameter, the positive and negative trial points are defined
as
\begin{equation}
\mathbf{p}_{j,\pm}^{(k)}
=
\mathbf{p}_{\mathrm{cur}}^{(k)}
\pm
\Delta_{j}^{(k)}\mathbf{e}_{j},
\qquad
j=1,\ldots,4 .
\label{eq:pattern_trials}
\end{equation}
where \(\mathbf{e}_{j}\) is the \(j\)-th Cartesian unit vector. Trial parameters are restricted to their prescribed admissible bounds. The
lower and upper bounds of the scaled optimization vector are defined as
\begin{equation}
\mathbf{p}_{\min}
=
\left[
0.10,\;
0.10,\;
0.50,\;
0.50
\right]^{\mathrm{T}},
\qquad
\mathbf{p}_{\max}
=
\left[
10,\;
10,\;
10,\;
10
\right]^{\mathrm{T}}.
\label{eq:parameter_bounds}
\end{equation}
In the present calculations, the initial step-size vector is
\begin{equation}
\bm{\Delta}^{(0)}
=
\left[
0.20,\;
0.20,\;
0.10,\;
0.10
\right]^{\mathrm{T}},
\label{eq:initial_step_size}
\end{equation}
consistently with the scaled optimization variables introduced in
Eq.~\eqref{eq:opt_parameters}.
For each parameter, the positive trial point is evaluated first. If its \(J\)
is lower than that of the current state, the trial point is accepted and
becomes the current point for the exploration of the next parameter.
Otherwise, the corresponding negative trial point is evaluated. If neither
trial point reduces the objective function, the current parameter vector is
retained for that coordinate.
If no trial point is accepted during a complete exploration of the four
parameters, the step-size vector is reduced according to
\begin{equation}
\bm{\Delta}^{(k+1)}
=
\gamma\,\bm{\Delta}^{(k)},
\qquad
0<\gamma<1.
\label{eq:pattern_shrink}
\end{equation}
Otherwise, the current step-size vector is retained for the next outer
iteration.

The pattern-search stage is terminated when either
\begin{equation}
\left\|
\bm{\Delta}^{(k)}
\right\|_{\infty}
\leq
\Delta_{\mathrm{tol}},
\label{eq:pattern_stop}
\end{equation}
or the prescribed maximum number of outer iterations is reached. In the
present calculations,
\(\gamma=0.5\),
\(\Delta_{\mathrm{tol}}=0.01\),
and the maximum number of outer pattern-search iterations is six.
The best parameter vector obtained during this stage is subsequently used
as the initial point for the quasi-Newton refinement.

Once the first-stage stopping condition is met, the best parameter set obtained
from the pattern-search phase is used as the starting point for a local
quasi-Newton refinement. The objective function remains the same full-field
\(J\) defined in Eq.~\eqref{eq:nrmse}. At the $\ell$-th quasi-Newton iteration,
the gradient of the objective function is approximated by forward finite
differences,
\begin{equation}
\left[\nabla J(\mathbf{p}^{(\ell)})\right]_i
\simeq
\frac{
J(\mathbf{p}^{(\ell)}+h_i\mathbf{e}_i)
-
J(\mathbf{p}^{(\ell)})
}{h_i},
\qquad i=1,\ldots,4,
\label{eq:qn_gradient}
\end{equation}
where \(J\) denotes the full-field NRMSE objective function, $h_i$ is the finite-difference
perturbation of the $i$-th parameter, and $\mathbf{e}_i$ is the corresponding
Cartesian unit vector.

The local search direction is then computed as
\begin{equation}
\mathbf{w}^{(\ell)}
=
-\mathbf{H}_{\ell}^{-1}
\nabla J(\mathbf{p}^{(\ell)}),
\label{...}
\end{equation}
where $\mathbf{H}_{\ell}^{-1}$ is an approximation of the inverse Hessian. A
damped trial update is generated as
\begin{equation}
\mathbf{p}_{\mathrm{trial}}^{(\ell)}
=
\mathbf{p}^{(\ell)}
+
\lambda\,\mathbf{w}^{(\ell)},
\label{eq:qn_damped_step}
\end{equation}
with $\lambda$ selected from the prescribed list of damping factors. The trial
state is accepted only if it reduces \(J\); otherwise, a smaller damping
factor is tested.
After an accepted update, the inverse-Hessian approximation is updated using a
BFGS-type formula. Defining
\begin{equation}
\mathbf{s}_{\ell}
=
\mathbf{p}^{(\ell+1)}-\mathbf{p}^{(\ell)},
\qquad
\mathbf{y}_{\ell}
=
\nabla J(\mathbf{p}^{(\ell+1)})
-
\nabla J(\mathbf{p}^{(\ell)}),
\label{eq:bfgs_sy}
\end{equation}
the update is written as
\begin{equation}
\mathbf{H}_{\ell+1}^{-1}
=
\left(\mathbf{I}-\rho_{\ell}\mathbf{s}_{\ell}\mathbf{y}_{\ell}^{\mathrm T}\right)
\mathbf{H}_{\ell}^{-1}
\left(\mathbf{I}-\rho_{\ell}\mathbf{y}_{\ell}\mathbf{s}_{\ell}^{\mathrm T}\right)
+
\rho_{\ell}\mathbf{s}_{\ell}\mathbf{s}_{\ell}^{\mathrm T},
\qquad
\rho_{\ell}
=
\frac{1}{\mathbf{y}_{\ell}^{\mathrm T}\mathbf{s}_{\ell}}.
\label{eq:bfgs_update}
\end{equation}
This update is applied only when the curvature condition is numerically
well-defined. Otherwise, the inverse-Hessian approximation is left unchanged.
In the present calculations, the finite-difference perturbation was
$h_i=0.01$ for all four scaled parameters, and the damping factors were
$\lambda=\{1.0,0.5,0.25,0.10,0.05\}$. 
The quasi-Newton stage was run for a fixed maximum of six iterations. Gradient-norm and parameter-update tolerances of
$10^{-3}$ were also monitored as potential early-stopping conditions, but in the present calculations the stage was terminated by reaching the six-iteration limit.

In the present study, the complete identification procedure required 128 evaluations of the
MATLAB--COMSOL workflow. The convergence plots, however, report only the accepted current
updates, i.e. the states that define the effective evolution of the identified parameters. In
particular, Figures~\ref{fig:nrmse_vs_evaluation} and \ref{fig:stiffness_eta & tau_vs_evaluation} show the evolution over these
accepted updates only. Therefore, the
neutral-axis deviation is used only to initialize the parameter set, whereas the final
calibration is governed by the minimization of the full-field discrepancy between the numerical
and experimental \(G_{11}\) distributions, measured by $J$.

\subsubsection{Evolution of the objective function and identified parameters}

\begin{figure}[t]
    \centering
    \includegraphics[width=0.55\textwidth]{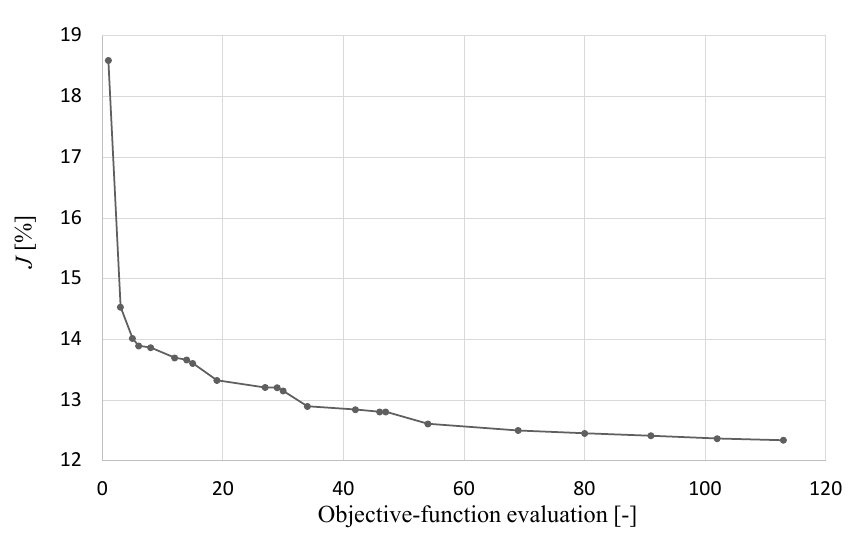}
    \caption{ Evolution of the objective function \(J\) over the accepted updates
of the identification procedure. The reported points correspond only to
accepted current states, i.e., to the effective optimization steps retained in
the parameter history. The complete identification required 128
objective-function calls of the coupled MATLAB--COMSOL workflow. The first
47 calls were associated with the pattern-search stage, whereas the remaining
calls were performed during the quasi-Newton refinement stage.}
    \label{fig:nrmse_vs_evaluation}
\end{figure}

The evolution of the objective function during the accepted optimization updates is reported
in Figure~\ref{fig:nrmse_vs_evaluation}. A marked reduction is observed in the initial part of
the procedure, followed by a progressively smoother decrease as the optimization approaches a
stable region of the parameter space. This trend indicates that the largest improvement is
achieved during the early search phase, while the subsequent updates mainly act as local
corrections that further reduce the discrepancy between the numerical and experimental
\(G_{11}\) fields.

\begin{figure}[H]
    \centering
    \includegraphics[width=0.95\textwidth]{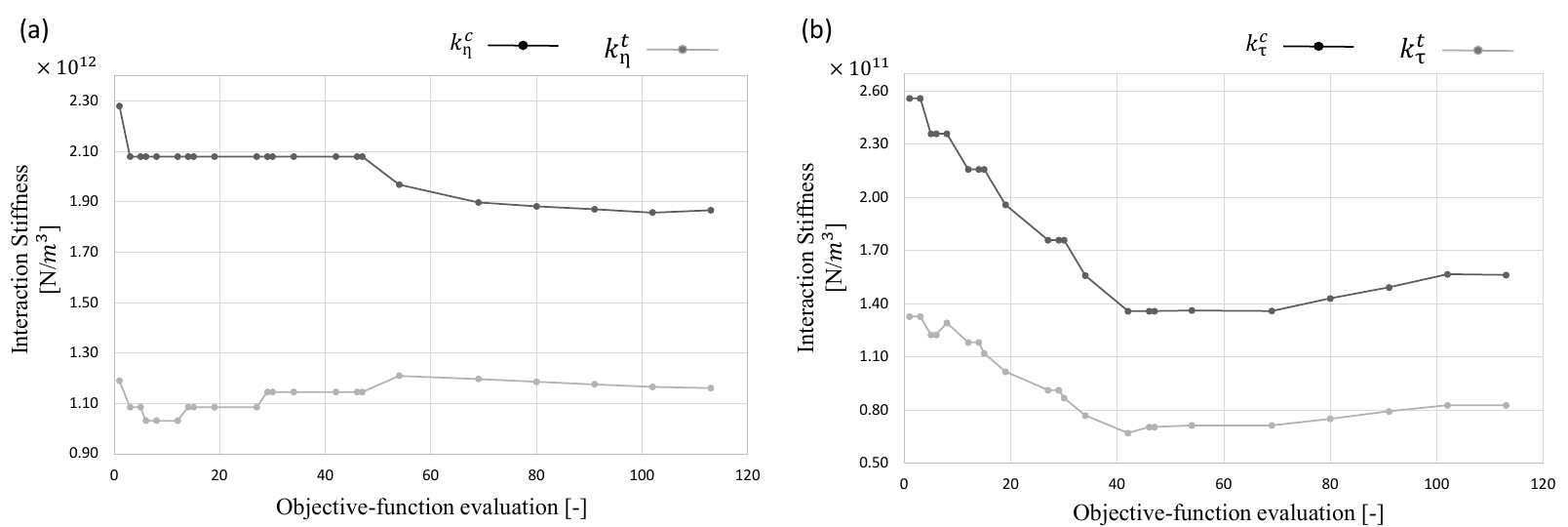}
    \caption{Evolution of the interaction stiffnesses retained during the optimization procedure:
(a) normal stiffnesses $k_{\eta}^{c}$ and $k_{\eta}^{t}$;
(b) tangent stiffnesses $k_{\tau}^{c}$ and $k_{\tau}^{t}$.
The horizontal axis denotes the objective-function evaluation number.
Only accepted parameter states are shown; therefore, the reported curves
represent the effective parameter history retained by the optimization procedure.}
    \label{fig:stiffness_eta & tau_vs_evaluation}
\end{figure}

The evolution of the identified normal and tangent  stiffnesses is reported in
Figure~\ref{fig:stiffness_eta & tau_vs_evaluation} (a) and (b),
respectively. These plots show only the accepted current updates and therefore represent the
effective parameter path retained by the algorithm. The tangent stiffnesses exhibit a more
pronounced relative variation during the optimization, whereas the normal stiffnesses evolve
more gradually. Overall, the combined reading of Figures~\ref{fig:nrmse_vs_evaluation} and \ref{fig:stiffness_eta & tau_vs_evaluation}
shows that the reduction of \(J\) is associated with a coordinated update of all four
identified parameters, rather than with the adjustment of a single stiffness component alone.

The final identified physical parameter set, \(\mathbf{q}^{\mathrm{opt}}\),
corresponding to the optimization process is summarized in
Table~\ref{tab:identification_steps}. Together with the convergence plots,
this table provides a compact representation of the parameter evolution
from the initial estimate to the final identified state.

\subsection{Comparison of the strain fields before and after optimization}

\begin{figure}[H]
    \centering
    \includegraphics[width=0.97\textwidth]{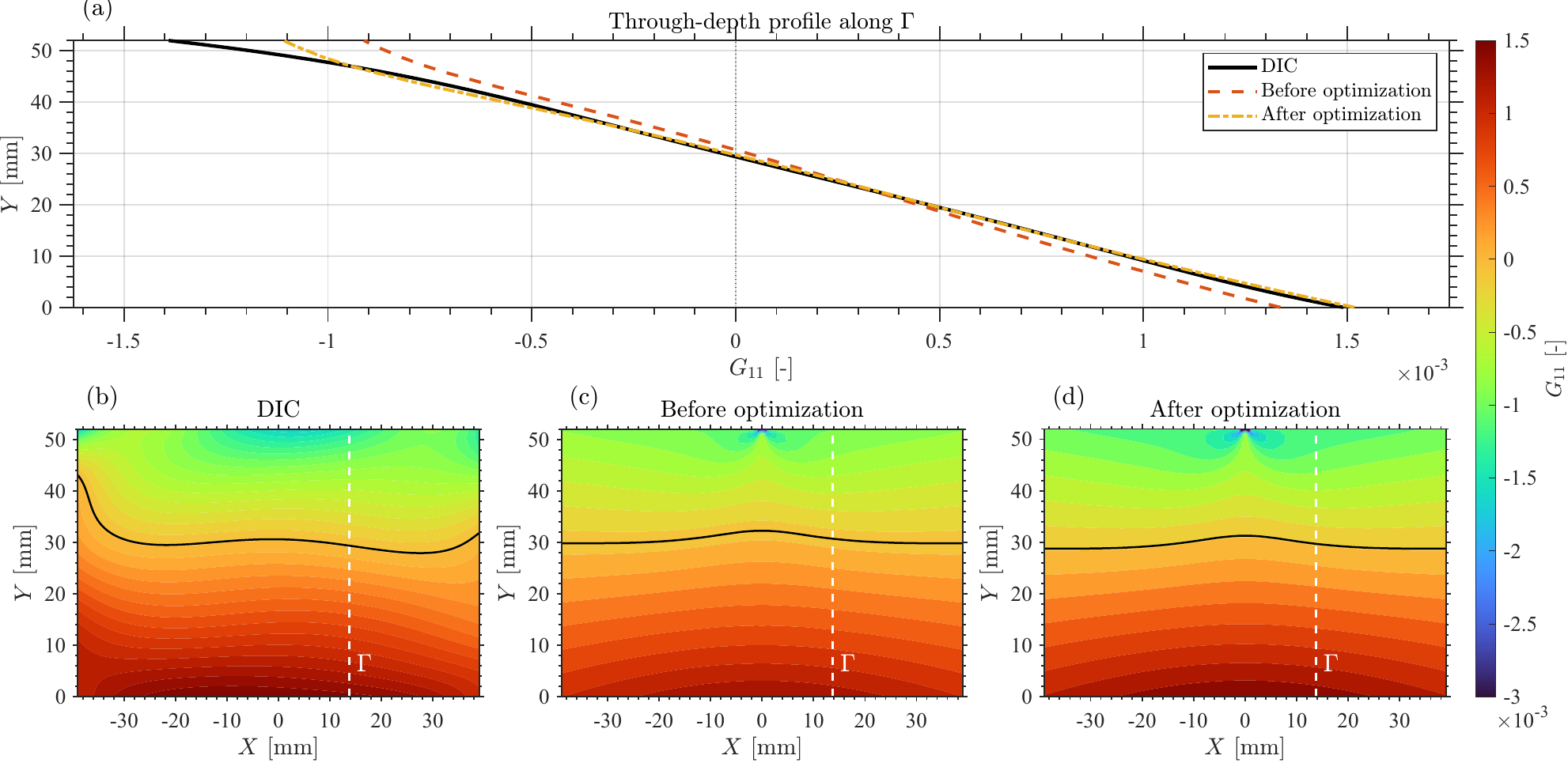}
    \caption{Comparison between experimental and numerical normal strain fields. Panel (a) reports the through-depth distributions of $G_{11}$ along the vertical profile $\Gamma$ for the DIC measurement, the numerical solution obtained from the initial parameter set, and the numerical solution after optimization. Panels (b)--(d) show the corresponding full-field distributions over the same ROI: (b) DIC measurement, (c) numerical result before optimization, and (d) numerical result after optimization. The ROI and the location of the profile $\Gamma$ are defined according to Figure~\ref{fig:beam_axes}(a). A common color scale is used for all contour plots to allow a direct comparison of the strain-field distributions. The solid black curves indicate the zero-strain line, corresponding to the neutral-axis position in the strain field, while the dashed vertical lines denote the profile $\Gamma$ used in panel (a).}
    \label{fig:step2_comparison}
\end{figure}

A direct comparison between the experimental field and the numerical predictions before
and after optimization is provided in Figure~\ref{fig:step2_comparison}. Panel (a)
compares the through-depth $G_{11}$ profiles along $\Gamma$, while panels (b)--(d) show the
corresponding full-field distributions over the ROI for the DIC measurement, the initial
numerical solution, and the optimized numerical solution, respectively. The comparison shows
that the initial parameter set already captures the overall qualitative structure of the normal
strain field, including the sign change across the depth and the general position of the neutral
axis. However, noticeable discrepancies remain both in the through-depth strain profile and
in the full-field contour levels. After optimization, the agreement improves significantly:
the numerical strain profile along $\Gamma$ becomes more consistent with the experimental one,
and the full-field distribution over the ROI better reproduces the observed strain gradients
and contour pattern.

It should be noted that the experimentally determined value
$\overline{Y}_{\mathrm{NA}}=30.20~\mathrm{mm}$ used in the initialization
represents the average neutral-axis position evaluated over multiple sections
within the DIC region, rather than the neutral-axis position at a single
section. The optimized numerical solution gives an average neutral-axis
position of $\overline{Y}_{\mathrm{NA}}=29.53~\mathrm{mm}$ over the same
region. At the specific section $\Gamma$
used for the through-depth comparison in
Figure~\ref{fig:step2_comparison}(a), the experimental neutral-axis position
is $Y_{\mathrm{NA}}(X_\Gamma)=29.31~\mathrm{mm}$, while the optimized numerical solution
predicts $Y_{\mathrm{NA}}(X_\Gamma)=29.65~\mathrm{mm}$. Therefore, although the average experimental neutral-axis
position was used only to provide the physically informed initialization of
the bimodulus ratio, the final full-field optimization retains a close
prediction of the experimentally observed neutral-axis position.

This improvement is also confirmed by the evolution of the objective function. In the present
case, \(J\) decreases from \(18.6\%\) for the initial parameter set to \(12.3\%\) after
optimization, indicating a substantial reduction in the discrepancy between the numerical and
experimental \(G_{11}\) fields. The remaining error is not uniformly distributed over the ROI.
A significant part of the mismatch is concentrated in the neighborhood of the load-application
region, where the comparison in Figure~\ref{fig:step2_comparison} reveals the largest local
differences between the measured and simulated fields.

From a mechanical viewpoint, these localized discrepancies may be associated with the highly concentrated nature of the applied load and with local effects that are not fully captured by the present first-gradient model. In particular, stress concentrations and localized load-transfer mechanisms in the vicinity of the loading nose are expected to affect the strain field over a small region below the applied force, where the experimental response departs more markedly from the idealized continuum prediction. 
Although the current identification procedure already provides a clear improvement in the global full-field agreement, a more refined description of the loading-zone behavior could be achieved in future developments by extending the present formulation toward strain-gradient continuum settings, which provide a natural framework for incorporating higher-order kinematics and microstructural regularization effects. 
At the same time, related generalized continuum approaches, including nonlocal, micropolar, and beam–continuum enriched formulations, may offer complementary descriptions of localized stress-transfer effects and enriched mechanical behavior near concentrated loading regions
\cite{eremeyev2020weak, ulukhanyan2025embedding, altenbach2015constitutive}.

Beyond the improvement in the local and global agreement between numerical and experimental
strain fields, these results also support the broader objective of the present study, namely the
identification of constitutive parameters from full-field experimental information. The
combined use of DIC measurements, the granular micromechanics-based numerical model, and
the proposed optimization framework allows one to determine a constitutive parameter set that
reproduces the measured bending response with good consistency. Therefore, the present
procedure suggests that full-field \(G_{11}\) data can be effectively used as a basis for the
inverse identification of the stiffness parameters entering the constitutive model.

Overall, the comparison indicates that the proposed identification procedure is able not only
to reduce the discrepancy between numerical and experimental strain fields, but also to
provide a constitutively meaningful parameter set consistent with the measured bending
response. In this sense, the final identified parameters offer a more reliable basis for
describing the material behavior than the DIC-informed initial estimate based on the neutral-axis shift.
\begingroup

\subsection{Parameter-resolution assessment}
\label{sec:parameter_resolution}

In the bending-dominated region considered in the identification, the
through-depth distribution of the longitudinal strain is approximately linear
at a given beam section and may be written as
\begin{equation}
G_{11}(X,Y)
\simeq
\kappa(X)\left[Y-Y_{\mathrm{NA}}(X)\right],
\label{eq:G11_bending_form}
\end{equation}
where \(\kappa(X)\) denotes the local curvature and
\(Y_{\mathrm{NA}}(X)\) the corresponding neutral-axis position. Although a
single through-depth profile may therefore be characterized mainly by its
slope and neutral-axis position, the inverse identification is not based on
one section. The objective function uses the complete two-dimensional
\(G_{11}\) field over the ROI, so that strain profiles at multiple
longitudinal positions are considered simultaneously. Since the bending
moment and curvature vary along the beam, these profiles provide a
spatially distributed set of observations rather than only two scalar
measurements.

To assess the resolution of the identified parameters relative to the
experimental uncertainty, the full-field discrepancy measure
\(\widetilde{J}\) is defined as
\begin{equation}
\widetilde{J}(\mathbf{p})
=
\sqrt{
\dfrac{1}{m}
\sum_{a=1}^{m}
\left[
G_{11}^{\mathrm{mod}}(a;\mathbf{p})
-
G_{11}^{\mathrm{obs}}(a)
\right]^2
}.
\label{eq:Jtilde}
\end{equation}
Since the denominator in the objective function \(J\) defined in
Eq.~\eqref{eq:nrmse} is independent of \(\mathbf{p}\), minimizing \(J\) is
equivalent to minimizing \(\widetilde{J}\). Therefore, both measures have
the same minimizing parameter vector, \(\mathbf{p}^{\mathrm{opt}}\). The dimensionless vector \(\mathbf{p}^{\mathrm{opt}}\) is obtained from the
final identified physical parameter set \(\mathbf{q}^{\mathrm{opt}}\) reported
in Table~\ref{tab:identification_steps}, according to the scaling and
stiffness-ratio definitions given in Eq.~\eqref{eq:opt_parameters}.

A one-at-a-time parameter analysis was then performed around
\(\mathbf{p}^{\mathrm{opt}}\). For each component \(p_i\), the perturbed
value was defined as
\begin{equation}
p_i(\delta)
=
p_i^{\mathrm{opt}}(1+\delta),
\label{eq:resolution_perturbation}
\end{equation}
while the remaining three components were kept fixed at their identified
values. The relative perturbation
\(\delta=\Delta p_i/p_i^{\mathrm{opt}}\) was prescribed at
\(-25\%\), \(-20\%\), \(-15\%\), \(-10\%\), \(-5\%\), \(0\%\),
\(5\%\), \(10\%\), \(15\%\), \(20\%\), and \(25\%\).
All simulations were evaluated using the same set of
\(m\) valid DIC points.

To express the experimental uncertainty on the same strain scale as
\(\widetilde{J}\), the no-load DIC measurements are used as a reference for
the measurement noise. In the absence of mechanical loading, the reference
strain is zero. For each no-load correlation, the magnitude of the
ROI-averaged measured \(G_{11}\) therefore provides a scalar measure of the
global no-load deviation. The largest absolute value among the four no-load
DIC correlations reported in Table~\ref{tab:noload_dic} is used to define
\begin{equation}
\widetilde{J}_{\mathrm{noise}}
=
\max_{k=1,\ldots,4}
\left|
\overline{G}_{11}^{(k)}
\right|.
\label{eq:Jtilde_noise}
\end{equation}
Using the values reported in Table~\ref{tab:noload_dic},
\(\max_k|\overline{G}_{11}^{(k)}|=3.66\times10^{-5}\).
At the identified parameter vector,
\(\widetilde{J}^{\mathrm{opt}}
=\widetilde{J}(\mathbf{p}^{\mathrm{opt}})
=9.47\times10^{-5}\).
The experimental resolution threshold
\(\widetilde{J}^{\mathrm{th}}\) is therefore
\begin{equation}
\widetilde{J}^{\mathrm{th}}
=
\widetilde{J}^{\mathrm{opt}}
+
\widetilde{J}_{\mathrm{noise}}
\label{eq:Jtilde_threshold}
\end{equation}

\begin{figure}[htb] \centering \includegraphics[width=0.92\textwidth] {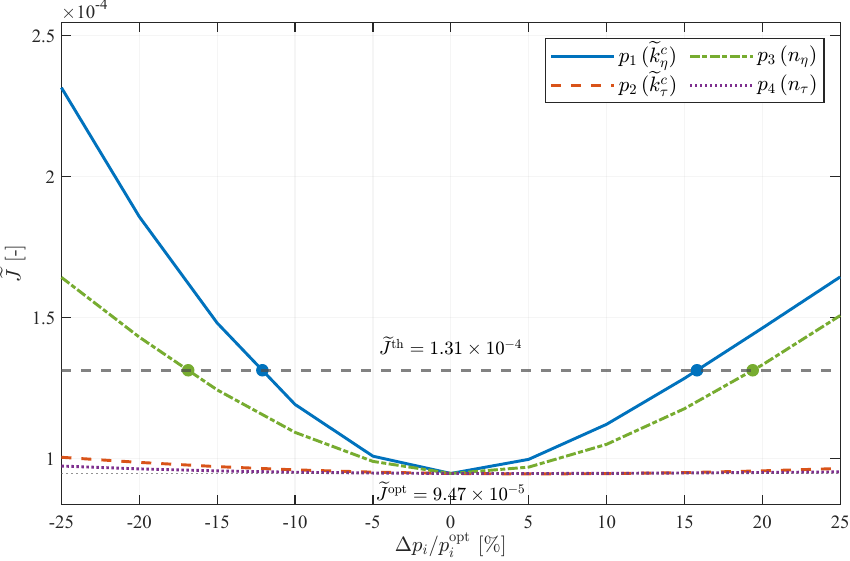} \caption{Variation of the discrepancy measure \(\widetilde{J}\) with independent relative perturbations of the four components of the dimensionless optimization vector \(\mathbf{p}^{\mathrm{opt}}\). The horizontal axis represents \(\Delta p_i/p_i^{\mathrm{opt}}\) in percent, while the vertical axis reports the dimensionless discrepancy measure \(\widetilde{J}\). The dotted horizontal line indicates \(\widetilde{J}^{\mathrm{opt}}=9.47\times10^{-5}\), and the dashed line denotes the resolution threshold \(\widetilde{J}^{\mathrm{th}} =\widetilde{J}^{\mathrm{opt}} +\widetilde{J}_{\mathrm{noise}} =1.31\times10^{-4}\), with \(\widetilde{J}_{\mathrm{noise}}=3.660\times10^{-5}\). The marked crossings occur at \(-12.10\%\) and \(+15.81\%\) for \(\widetilde{k}_{\eta}^{c}\), and at \(-16.86\%\) and \(+19.39\%\) for \(n_{\eta}\). No crossing is observed for \(\widetilde{k}_{\tau}^{c}\) or \(n_{\tau}\) within the investigated \(\pm25\%\) interval.}
\label{fig:parameter_resolution_noise} \end{figure}

Figure~\ref{fig:parameter_resolution_noise} shows the variation of
\(\widetilde{J}\) with the independent relative perturbation of each
component of \(\mathbf{p}^{\mathrm{opt}}\).
Linear interpolation between adjacent evaluated perturbation levels gives
threshold crossings at \(-12.10\%\) and \(+15.81\%\) for
\(\widetilde{k}_{\eta}^{c}\), and at \(-16.86\%\) and \(+19.39\%\)
for \(n_{\eta}\). Neither \(\widetilde{k}_{\tau}^{c}\) nor \(n_{\tau}\)
reaches the threshold within the investigated \(\pm25\%\) interval.

These results show that the four components of the constitutive
parameterization are not constrained with equal experimental resolution by
the present three-point bending test. The parameters associated with the
normal interaction, \(\widetilde{k}_{\eta}^{c}\) and \(n_{\eta}\), are
substantially better resolved, whereas the measured \(G_{11}\) field is
considerably less sensitive to the parameters associated with the tangent
interaction.

This difference is also related to the experimental quantity used in the
identification. The objective function is constructed exclusively from the
measured \(G_{11}\) field. Although the normal strain in the \(Y\)-direction,
\(G_{22}\), was also obtained from the DIC measurements, its magnitude is
considerably smaller than that of \(G_{11}\), while the measurement noise is
of the same order. Consequently, \(G_{22}\) has a lower signal-to-noise ratio
and does not provide a sufficiently accurate basis for the present
identification.
The calibration is therefore mainly driven by \(G_{11}\), which, for the
present bending configuration, provides limited sensitivity to the tangent
interaction parameters.

The four-parameter formulation is retained because the objective of the
present work is to identify simultaneously the complete set of independent
parameters governing the bimodulus interaction response. The resolution
analysis shows, however, that these parameters are not constrained with equal
strength by the present experiment. Accordingly, the identified parameters
should be interpreted collectively as a parameter set reproducing the
measured full-field bending response, rather than as four independently and
equally resolved material constants. Future experiments providing a more
accurate measurement of \(G_{22}\), with an improved signal-to-noise ratio,
could provide complementary information for the identification of the
tangent interaction parameters.

The effect of the finite-difference increment was further examined to
assess whether the value \(h_i=0.01\) used in the quasi-Newton refinement
is consistent with the experimental resolution of the DIC measurements.
This check is relevant because the gradients used by the quasi-Newton
procedure are evaluated from finite differences of the numerical response.
For \(h_i=0.01\), the symmetric finite-difference response of the numerical
\(G_{11}\) field was quantified as
\begin{equation}
\Delta G_{11,i}=
\frac{1}{2}
\sqrt{\frac{1}{m}\sum_{a=1}^{m}
\left[
G_{11}^{\mathrm{mod}}(a;\mathbf{p}^{\mathrm{opt}}+h_i\mathbf{e}_i)
-
G_{11}^{\mathrm{mod}}(a;\mathbf{p}^{\mathrm{opt}}-h_i\mathbf{e}_i)
\right]^2}.
\end{equation}
The resulting values are
\(1.07\times10^{-4}\), \(1.77\times10^{-5}\),
\(8.05\times10^{-5}\), and \(1.03\times10^{-5}\) for
\(\widetilde{k}_{\eta}^{c}\), \(\widetilde{k}_{\tau}^{c}\),
\(n_{\eta}\), and \(n_{\tau}\), respectively. These values can be
compared with the adopted no-load DIC noise level of
\(3.66\times10^{-5}\). The response changes associated with
\(\widetilde{k}_{\eta}^{c}\) and \(n_{\eta}\) are therefore above the
noise level, whereas those associated with
\(\widetilde{k}_{\tau}^{c}\) and \(n_{\tau}\) are below it. Nevertheless,
\(h_i=0.01\) is retained for the quasi-Newton refinement because the
optimization is performed on the complete deterministic objective
constructed from the fixed experimental field. The finite-difference
increment is used to evaluate the local numerical gradients and is not
interpreted as an experimentally resolved parameter variation. This
result is also consistent with the parameter-resolution analysis, which
shows that the tangent-interaction parameters are less strongly resolved
by the present \(G_{11}\)-based bending experiment.

\section{Conclusions}

The present study proposed an identification procedure for the stiffness parameters of a
bimodulus constitutive model formulated within a granular micromechanics framework,
combining full-field DIC measurements from three-point bending tests with a hybrid
numerical optimization strategy. Starting from an initial estimate based on the experimental
neutral-axis shift, the parameter set was refined through a direct comparison between the
experimental and numerical \(G_{11}\) fields over the selected ROI, using \(J\) (NRMSE) as
objective function.

The obtained results show that the combined use of DIC, forward numerical simulations,
and hybrid optimization makes it possible to identify a constitutive parameter set that
reproduces the measured bending response with good consistency. In particular, the procedure
led to a clear reduction of the discrepancy between numerical and experimental full-field
strain data, improving both the through-depth profile along \(\Gamma\) and the spatial
distribution of \(G_{11}\) over the ROI. In this sense, the proposed framework supports the
use of full-field experimental information as a basis for the inverse identification of the
stiffness parameters entering the constitutive model.  The parameter-resolution analysis further shows that the four identified
parameters are not constrained with equal strength by the present bending
experiment, with the normal-interaction parameters being more strongly
resolved than the tangent-interaction parameters.

At the same time, the present study also highlights the current limitations of the procedure.
Although the DIC analysis provides the in-plane strain components \(G_{11}\), \(G_{22}\), and
\(G_{12}\), only \(G_{11}\) has been  retained in the identification process. The reason is that the
magnitudes of \(G_{22}\) and \(G_{12}\) are significantly smaller and therefore much more
sensitive to measurement noise, so that they could not provide a sufficiently robust basis for
the present inverse analysis. Consequently, the identification was deliberately constructed on
the dominant normal strain field \(G_{11}\) only.

Another important point is that both the experimental interpretation and the numerical
model have been restricted to the elastic regime. The experimental campaign has been designed so as to
keep the material response, as far as possible, within an admissible elastic range, while the
forward model adopted in this work is purely elastic and does not include plasticity or other
inelastic effects. For this reason, the present identification should be understood as an
elastic constitutive calibration based on full-field bending strains.

The remaining discrepancy after optimization is mainly localized in the neighborhood of the
load-application region. This localization is expected and suggests that part of the residual error may be associated with
strongly localized effects induced by the localized load-transfer mechanism and by the idealized representation
of the contact region, which are not fully captured by
the present first-gradient formulation. In this respect, a natural continuation of the present
work is the extension toward a strain-gradient version of the model, which may provide a more
refined description of local stress-transfer mechanisms and reduce the mismatch in the
vicinity of the loading point. Such an extension is also relevant from the
viewpoint of parameter identification. In the present first-gradient formulation,
the characteristic length \(L\) is prescribed, and the reported stiffness
coefficients should therefore be interpreted as conditional on this prescribed
length scale. Indeed, an equivalent first-gradient response would require a
\(1/L^2\) rescaling of the interaction stiffnesses if \(L\) were changed. In a
strain-gradient formulation, \(L\) would contribute to higher-order terms and
should therefore be identified as an additional parameter together with the four independent bimodulus parameters considered in this work.

Another possible development of the present framework concerns the post-processing
interpretation of the identified granular stiffnesses in terms of equivalent macroscopic elastic
parameters under idealized uninormal loading conditions, consistently with Eq.~(47) of~\cite{misra2025emergence}. In this way, the corresponding
tensile and compressive Young's moduli and Poisson's ratios could be derived, thus providing
a more direct link between the identified granular model and the conventional engineering
constants commonly reported in the literature.

Overall, the proposed procedure represents a first step toward the identification of
constitutive parameters in bimodulus granular micromechanics directly from full-field
experimental measurements. While the present study is limited to the elastic range and to
the use of the dominant \(G_{11}\) strain component, the results indicate that the approach is
promising and can be further extended in future work toward richer kinematics, improved
local modeling, and more comprehensive identification strategies.
% =================================================

\bibliographystyle{elsarticle-num} 
\bibliography{bibliography}

\end{document}